\documentclass[authoryear,3p,11pt]{elsarticle}

\usepackage{amsmath,amssymb,mathtools,mathrsfs,structmech,mathpazo,float}
\usepackage{cleveref}
\usepackage{ulem}
\usepackage{soul}
\usepackage{bm}
\usepackage{cancel}
\usepackage{subcaption}
\DeclareUnicodeCharacter{2212}{-}
\newcommand{\beq}{\begin{equation}}
\newcommand{\eeq}{\end{equation}}

\newcommand{\upd}{\mathrm{d}}
\newcommand{\br}{\bar{r}}
\newcommand{\bc}{\bar{c}}
\newcommand{\bt}{\bar{t}}
\newcommand{\bu}{\bar{u}}
\newcommand{\bZ}{\bar{\zeta}}
\newcommand{\amax}{a_{\max}}

\newcommand{\bp}{\bar{p}}
\newcommand{\bOmega}{\tilde{\Omega}}
\newcommand{\bSigma}{\bar{\sigma}}
\newcommand{\bPsi}{\bar{\psi}}

\newcommand{\ETwoD}{E_{2D}}
\newcommand{\Rd}{R_{\mathrm{drum}}}
\newcommand{\Tpre}{T_{\mathrm{pre}}}

\newcommand{\kone}{\alpha_1}
\newcommand{\ktwo}{\alpha_2}
\newcommand{\kthree}{\alpha_3}
\newcommand{\cc}{\mathrm{c.c.}}
\newcommand{\ii}{\mathrm{i}}
\newcommand{\bending}{{\mathcal B}}

\journal{International Journal of Nonlinear Mechanics}

\begin{document}

\begin{frontmatter}

\title{Axisymmetric membrane nano-resonators: A comparison of nonlinear reduced-order models}

\author{Safvan Palathingal$^{\dag}$ and Dominic Vella$^{\ddag}$}

\address{$^{\dag}$\:Department of Mechanical and Aerospace Engineering, Indian Institute of Technology Hyderabad,Telengana, India\\
$^{\ddag}$\:Mathematical Institute, University of Oxford, Woodstock Rd, Oxford, OX2 6GG, UK}

\begin{abstract}
The shift in the backbone of the frequency--response curve and the `jump-down' observed at a critical frequency observed in nano-resonators are caused by their nonlinear mechanical response. The shift and jump-down point are therefore often used to infer the mechanical properties that underlie the nonlinear response, particularly the resonator's stretching modulus.  To facilitate this, the resonators's dynamics are often modelled using a Galerkin-type numerical approach or lumped ordinary differential equations like the Duffing equation, that incorporate an appropriate nonlinearity. To understand the source of the problem's nonlinearities, we first develop an axisymmetric but spatially-varying model of a membrane resonator subject to a uniform oscillatory load with linear damping. We then derive asymptotic solutions for the resulting partial differential equations (PDEs) using the Method of Multiple Scales (MS), which allows a systematic reduction to a Duffing-like equation with  analytically determined coefficients. We also solve the PDEs numerically via the method of lines. By comparing the numerical solutions with the asymptotic results, we demonstrate that the numerical approach reveals a non-constant maximum compliance with increasing load, which contradicts the predictions of the MS analysis. In contrast, we show that combining a Galerkin decomposition with the Harmonic Balance Method accurately captures the non-constant maximum compliance and reliably predicts  jump-down behaviour.  We analyze the resulting frequency-response predictions derived from these methods. We also argue that fitting based on the  jump-down point may be sensitive to noise and discuss strategies for fitting frequency-response curves from experimental data to theory that are robust to this.
\end{abstract}


\end{frontmatter}

%%%%%%%%%%%%%%%%%%%%%%%%%%%%%%%%%%%%%%%%%%%%%%%%%%%%%%
\section{Introduction}

Nano-resonators have recently emerged as a powerful means of sensing \cite[][]{Eom2011}, whether to detect the presence of single molecules \cite[][]{Naik2009} or to measure the stretching modulus of two-dimensional materials such as graphene \cite[][]{davidovikj_nonlinear_2017}. In each application, the common idea is that by measuring the amplitude of the system's response to an imposed high frequency oscillation, properties of the system (such as the inertia or restoring force) can be accurately measured.

A common experimental realization of a nano-resonator consists of an elastic beam or plate clamped above an electrode, but with a finite gap \cite[see][for example]{Pelesko2002}. Upon applying an AC potential difference, the elastic element vibrates and its maximum amplitude can be measured as a function of the driving frequency and the amplitude of the applied voltage. A number of different perspectives and techniques have been taken to model the relationship between the driving frequency and load and the oscillation amplitude that is observed \cite[see][for reviews]{LifshitzReview,Xu2022}.

Since the shape of the resonator plate evolves in time, the underlying equations are partial differential equations (PDEs) with some form of nonlinearity (either because of a coupling between stress and deformation or because of a nonlinear electrostatic forcing). These equations are normally solved numerically by using a Galerkin procedure in which the shape of the plate is resolved into a superposition of modes with time-varying amplitude. This approach allows for the PDE to be reduced to a series of coupled nonlinear ODEs that can be solved to determine the amplitude-frequency response \cite[see][for example]{Farokhi2016,Farokhi2023}. In this way complex material and geometrical properties may be modelled via reduced-order systems, as in, for example, the work on composite shells of \cite{mahmure2021primary}, \cite{sofiyev2021influences} and \cite{sofiyev2023nonlinear}.

While numerical approaches using a large number of Galerkin modes yield some insight, they are of limited utility in the interpretation of experiments. For example, a common application of the shift observed in the backbone of the frequency--response curve is to infer the nonlinear stiffness of an elastic membrane from the experimentally-observed frequency-amplitude response \cite[see][for example]{davidovikj_nonlinear_2017}, and especially to focus on the frequency at which the amplitude suddenly `jumps down'. A purely numerical approach to this problem is difficult because it is very computationally intensive --- one needs to solve the governing PDEs for many sets of parameter values. Hence simpler approaches are frequently proposed. One of these is to use a single term Galerkin representation, allowing analytical progress to be made to determine, for example, the frequency-amplitude response \cite[][]{Kacem2009}; this is often referred to as the Harmonic Balance method \cite[see][for example]{Kim2003}.  The Harmonic Balance Method works by assuming a Fourier series solution to the ODE derived from the reduced PDE, transforming the problem into a system of algebraic equations. The solution can often be expressed in closed form, depending on the number of terms used in the series approximation. This method is particularly effective for a range of nonlinear oscillations, especially for Duffing-like equations \cite[][]{Krack.2019}. For even strongly nonlinear systems, various modified versions of the Harmonic Balance Method are also employed to enhance accuracy, computational speed, and applicability \cite[][]{Yan.2023}.

Alternatively, `semi-analytical' methods may be used in which a Duffing type equation is posited, allowing the analysis to proceed; however, there remains the difficult question of how to relate the different coefficients of the posited model to the parameters of the physical problem of interest. Moreover, the Duffing equation has only a single degree of freedom (dof) while the continuous problem effectively has infinitely many dof. It is natural to wonder whether a single dof system can satisfactorily describe the full vibrations of a membrane. While many studies use these coefficients as fitting parameters \cite[][]{Mestrom2008,Chowdhury2020}, \cite{davidovikj_nonlinear_2017} used a power series description of the spatial variation of the  displacement to determine numerically the effective mass in the Duffing model as a fraction of the true mass of the membrane, as well as the effective cubic stiffness. Even so, the effective damping coefficient remained as a fitting parameter here, via the quality factor. Moreover, the numerical calculation of, for example, the cubic stiffness needs to be repeated for membranes of different Poisson ratio. In this paper we revisit this question from an analytical perspective.

We begin with a systematic reduction of the governing PDEs to ODEs, and then aim to derive a fully analytical expression that captures the nonlinear characteristics of the amplitude-frequency response. This expression should be  accurate but sufficiently simple to allow for the inference of the stiffness of an elastic membrane by fitting it to experimentally observed frequency-amplitude data. While the Multiple Scales method is typically employed in such cases, the Harmonic Balance method is known to provide more accurate solutions for strongly nonlinear differential equations compared to perturbation methods \cite[][]{Lim2003, Lai2009, Gottlieb.2004}, especially for Duffing-like equations. Notably for the the Harmonic Balance method, even using a single term in the Fourier series yields a tractable analytical solution for the force-response curve  \cite[][]{Krack.2019,Mickens96}. In this work, we derive and compare the analytical expressions obtained from both methods with detailed numerical results based on a spatial discretization of the governing PDEs and the method of lines (rather than using a Galerkin decomposition). We  focus on nonlinear features of the numerical results (such as the shift of the backbone curve, the points of maximum compliance  and the point at which `jump-down' occurs) and the extent to which the different analytical techniques satisfactorily describe these features.

In terms of formal analysis, various papers, starting with \cite{Younis2003}, use the method of Multiple Scales, together with a solvability condition, to determine an amplitude equation for the oscillations of a two-dimensional beam subject to an alternating potential difference. Their work accounted for in-plane stretching and they were able to derive frequency-amplitude response curves for a range of forcing amplitudes. The analysis of \cite{Younis2003} was simplified slightly by the fact that the stress within a stretched beam is spatially uniform; this is not the case for a stretched axisymmetric plate. More recently, \cite{Caruntu2016} and \cite{Caruntu2023} used a similar technique to model oscillations of a clamped circular plate, but did not account for the possibility of mid-plane stretching, or indeed of any tension in the mid-plane of the plate. As a result, their results only apply to amplitudes of deformation that remain small compared to the thickness of the membrane \cite[][]{Chopin2008} --- an assumption that is not typical of many clamped plate resonators \cite[for example][observe vertical deformations of $\gtrsim5\mathrm{~nm}$]{davidovikj_nonlinear_2017}.

In this paper, we consider the problem of an elastic membrane (i.e.~we neglect the bending stiffness of the sheet). We begin in \S\ref{sec:Models} by developing our mathematical model of a membrane subject to a time-varying, but spatially uniform, load. Then, in \S\ref{sec:theory}, we derive asymptotic solutions to these PDEs using the Method of Multiple Scales. In \S\ref{sec:Numerics}, we numerically solve the PDEs using the method of lines and compare the results with analytical predictions from the Method of Multiple Scales. While the results generally agree for small amplitudes, there are significant discrepancies at larger amplitudes.  We therefore explore an alternative approach using a Galerkin method combined with Harmonic Balance in \S \ref{sec:HB} to better account for our numerical solutions. Moreover, our approach shows how the governing PDEs can be systematically reduced to a single ordinary differential equation, specifically the Duffing equation. Unusually, we directly (and analytically) relate the  parameters in the corresponding Duffing equation to the corresponding parameters in the model. We then show that the numerically-determined frequency--response curves are well described by the results of a Harmonic Balance analysis with no fitting parameters. In \S\ref{sec:fitting}, we then use the comparison between frequency--response curves, and the difference in jump-down points observed in particular, to suggest  a more robust method for fitting experimental data to these curves before summarizing our results in \S\ref{sec:Conclusion}.

\section{Modelling\label{sec:Models}}

\begin{figure}[htbp]
    \centering
\includegraphics[width=0.6\textwidth]{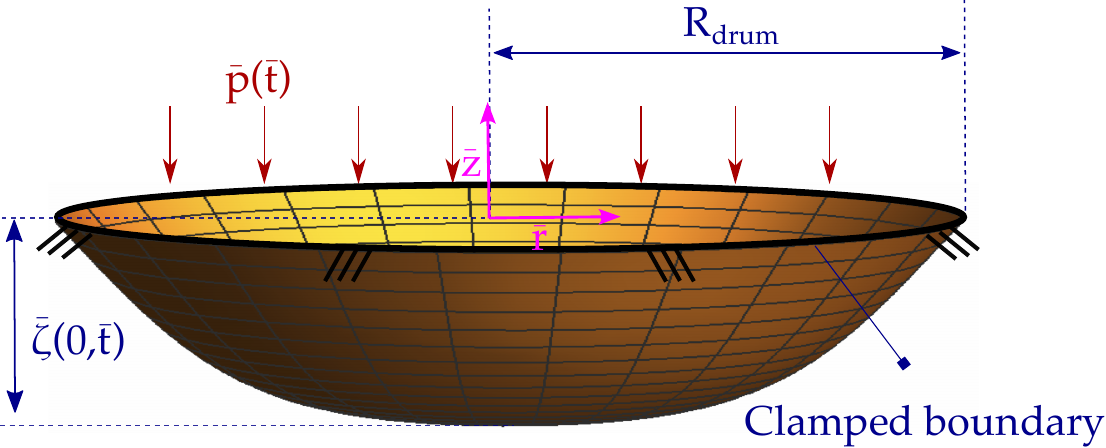}
    \caption{The axisymmetric problem considered here: A circularly clamped elastic membrane, under pre-tension $\Tpre$, is subject to an oscillatory, but spatially uniform, pressure $\bp(\bt)$. (Bars are used to denote dimensional quantities; dimensionless quantities do not carry bars.)}
    \label{fig:Setup}
\end{figure}

 We model the  nano-resonator as a thin (i.e.~two-dimensional),   elastic membrane with clamped circular boundary at a radius $\Rd$.  The membrane is subject to a uniform pre-tension, $\Tpre$, as well as a spatially uniform, but time-varying, pressure $\bp(\bt)$. We assume that the membrane has a mass per unit area $\rho h$ (so that $h$ is a nominal thickness and $\rho$ a nominal density). As a result of the oscillating applied pressure, the membrane is displaced transversely so that it adopts a profile, $\bZ(\br,\bt)$, that varies in both time and space. The setup is illustrated schematically in fig.~\ref{fig:Setup}.

\subsection{Governing equations}

 To simplify the problem, we make two key  assumptions: (i) the effects of the membrane's bending stiffness may be neglected and (ii) the stress profile within the membrane is instantaneously determined by its shape. The appropriateness of these assumptions for the typical space and time scales of nano-resonators is discussed in more detail in \ref{sec:SimplifyingAssumptions}. With these assumptions, we use the standard equations of vertical and in-plane force balance on an elastic membrane \cite[][]{nayfeh_nonlinear_2008}, presenting these first in dimensional form, before non-dimensionalizing appropriately.

\subsection{Dimensional problem}

The vertical force balance equation for a (zero bending stiffness) membrane subject to linear damping  is given \cite[see][for example]{nayfeh_nonlinear_2008} by
\begin{equation}
\rho h \frac{\partial^2 \bZ}{\partial \bt^2} + \bc h \frac{\partial \bZ}{\partial \bt}=\bp(\bt)+\bm{\nabla}\cdot\left( \bm{\bSigma}\cdot\bm{\nabla}\bZ \right),
    \label{Eq:governing_out_of_plane}
\end{equation} where   $\bc$ is the damping coefficient, $\bm{\bSigma}$ denotes the stress tensor and other quantities were defined in the text before fig.~\ref{fig:Setup}.  The boundary conditions on the transverse displacements are:
\begin{equation}
    \left.\frac{\partial\bZ}{\partial r}\right|_{r=0}=0,\quad \bZ(\Rd)=0.
    \label{Eq:out_of_plane_bcs}
\end{equation} Here the slope, $\partial\bZ/\partial \br$, is zero at the centre, $\br=0$, since we restrict our attention to axisymmetric solutions and there is no vertical displacement at the outer edge ($\br=\Rd$) where it is clamped. 

Assuming the in-plane force balance is satisfied instantaneously (no in-plane acceleration --- see  \ref{sec:SimplifyingAssumptions}) we also have
\begin{equation}
\bm{\nabla}.\bm{\bSigma}=0.
    \label{Eq:governing_in_plane}
\end{equation} To ensure that this equation is automatically satisfied in all that follows, we introduce the first integral of the Airy stress function, which we denote $\bPsi(\br,\bt)$ and is defined such that
  \begin{equation}
      \bSigma_{r r} =\frac{\bPsi}{\br},\quad  \sigma_{\theta \theta}=\frac{\partial \bPsi}{\partial \br}.
  \end{equation} The stress function $\bPsi$ is determined by the requirement that the corresponding displacement field is physically realizable --- the so-called compatibility equation
  \begin{equation}
      \br\frac{\partial}{\partial \br}\left[\frac{1}{\br}\frac{\partial(\br\bPsi)}{\partial \br}\right] =-\frac{\ETwoD}{2} \left(\frac{\partial\bZ}{\partial \br}\right)^2,
    \label{Eq:compatibility_dimensional}
  \end{equation} where  $\ETwoD = E h$, is the two-dimensional Elastic modulus of the membrane.  

While considering the boundary conditions for in-plane displacements $\mathbf{\bu}=\bu_{\bar{r}} (\br,\bt)\mathbf{e}_{\br}$, we need to account for the additional deformation ($\bu_0(\br)$) imposed prior to loading by the pre-tension, $\Tpre$. Though there is no deformation at the centre caused by pre-tension, there is a nonzero horizontal displacement ($\bu_0(\Rd)$) where the clamping is imposed.  This value remains unchanged when the sheet is later loaded with pressure. Hence we must have:  
\begin{equation}
    \bu_r(0,\bt)=0,\quad \bu_{\br}(\Rd,\bt)=(1-\nu)\frac{\Tpre\Rd}{\ETwoD}.
    \label{Eq:in_plane_bcs}
\end{equation} Using the Hookean constitutive relationship, these boundary conditions can be expressed in terms of the stress function as
\begin{equation}
    \lim_{\br\to0}\left[\br\frac{\partial\bPsi}{\partial \br}-\nu\bPsi\right]=0,\quad \Rd\left.\frac{\partial\bPsi}{\partial r}\right|_{\Rd}-\nu\bPsi(\Rd,\bt)=(1-\nu)\frac{\Tpre}{\ETwoD}.
    \label{eqn:BCs_Psi_Dim}
\end{equation}
 
\subsection{Non-dimensionalization}
\label{Sec:Non-dimensionalization}
To determine the dimensionless forms of \cref{Eq:governing_out_of_plane,Eq:compatibility_dimensional}, we introduce the dimensionless variables:
\begin{align}
    r=\frac{\br}{\Rd} , \quad 
    t=\bt  \left(\frac{\Tpre}{\rho h\Rd^2}\right)^{1/2} ,  \quad{\bm{\sigma}}= \bm{\bSigma}/\Tpre , \nonumber\\ {\zeta}({r},{t})=\bZ(\br,\bt)\left(\frac{\ETwoD}{\Tpre\Rd^2}\right)^\frac{1}{2} ,  
    %\quad \bar{u}=\frac{u}{\Rd}\frac{\ETwoD}{\Tpre},
    \quad {\psi}({r},{t})=\frac{\bPsi(\br,\bt)}{\Tpre\Rd}.\label{Eq:NDvariables}
\end{align}
The dimensionless version of \cref{Eq:governing_out_of_plane} is then
\begin{equation}
\frac{\partial^2 {\zeta}}{\partial {t}^2}+  \gamma \frac{\partial {\zeta}}{\partial {t}}={ p}({r},t)+\frac{1}{r}\frac{\partial}{\partial r}\left( {\psi}\frac{\partial{\zeta}}{\partial {r}} \right),
    %\label{Eq:governing_out_of_plane_dimensionless}
    \label{Eq:VForceBal_ND}
\end{equation}
 where \begin{equation}
{p}=\bp\frac{\Rd \ETwoD^{1/2}}{\Tpre^{3/2}},     \label{Eq:normal_p}     
 \end{equation} is the dimensionless forcing pressure and
\begin{equation}
    \gamma= \bc  \left(\frac{ h\Rd^2}{\rho  \Tpre}\right)^{1/2} ,\label{Eq:gamma}
\end{equation} is the dimensionless damping coefficient. The dimensionless  compatibility equation is given by
\begin{equation}
    {r}\frac{\partial}{\partial{r}}\left[\frac{1}{{r}}\frac{\partial({r}{\psi})}{\partial{r}}\right] =-\frac{1}{2} \left(\frac{\partial{\zeta}}{\partial{r}}\right)^2.
    \label{Eq:compatibility_dimensionless}
\end{equation}
 \Cref{Eq:VForceBal_ND,Eq:compatibility_dimensionless} are to be solved subject to the dimensionless boundary conditions:
\begin{equation}
    \left.\frac{\partial{\zeta}}{\partial {r}}\right|_{r=0}=0,\quad {\zeta}(1,t)=0, \quad 
    \lim_{{r}\to0}\left[{r}\frac{\partial{\zeta}}{\partial {r}}-\nu{\psi}\right]=0,\quad \left.\frac{\partial{\psi}}{\partial {r}}\right|_{r=1}-\nu{\psi}(1,t)=(1-\nu)
    \label{Eq:bcs_dimensionless}.
\end{equation}
In our numerics, we use a purely oscillatory forcing pressure,
\begin{equation}
 p=P\cos (\Omega t+\phi),
 \label{eqn:PressureImposed}
\end{equation} where $\phi$ is a phase. For simplicity, the pressure in \eqref{eqn:PressureImposed} is spatially uniform and has zero time-average. However, in many experiments a DC bias is applied in addition to an AC voltage \cite[see][for example]{davidovikj_nonlinear_2017}.

Our primary goal is to determine the oscillation amplitude expected at a given frequency --- the so-called frequency-amplitude response curve \cite[][]{nayfeh_nonlinear_2008}. There are two common asymptotic techniques used to determine this: Multiple Scales \cite[][]{LifshitzReview} and Harmonic Balance \cite[][]{nayfeh_nonlinear_2008}. We begin by considering the method of Multiple Scales, which is a systematic approach but, as we shall see, leads to some discrepancies with numerical results. The key difficulty here (in comparison to the standard Duffing equation) is to account for the spatial, as well as temporal, variation in the membrane's deflection and stress profiles.

%%%%%%%%%%%%%%%%%%%%%%%%%%%%%%%%%%%%%%%%%

\section{Theory: Multiple Scales approach\label{sec:theory}}

We now seek to determine asymptotic results for the amplitude of oscillations as a function of the forcing pressure and frequency (as well as the linear damping).

\subsection{Expansion and scalings}

In the Multiple Scales approach, we assume that the motion takes place over a time scale of order unity (comparable to the frequency of the natural, small displacement oscillation) but also that the oscillation amplitude itself evolves over a longer, slow, time scale. We therefore introduce two time scales, $t_0=t$ and $t_1=\epsilon t$,  where $\Omega=\omega_0+\epsilon\varsigma$ and $\epsilon\ll1$ --- here $\varsigma$ is the detuning parameter and measures how far from the resonant frequency, $\omega_0$, the oscillatory forcing is.  With these two time scales, we  note that
\begin{equation}
    \frac{\partial^2}{\partial t^2}=\frac{\partial^2}{\partial t_0^2}+2\epsilon \frac{\partial^2}{\partial t_0\partial t_1}+O(\epsilon^2).
    \label{eqn:TimeDerivative}
\end{equation}

Our aim is to understand how the frequency at which the jump-down is observed is affected by the change in the stress within the membrane caused by deformation. We have defined $\epsilon$ to be the scale of the shift in the maximum response frequency that is caused by the modification to the stress and so we expect that this perturbation to the stress enters also at $O(\epsilon)$; we therefore let
\begin{equation}
\psi(r,t_0,t_1)=r+\epsilon\psi_2(r,t_0,t_1)+...
\end{equation} where the leading order (linear in $r$) term represents the state of uniform, isotropic pre-tension that exists prior to any vibration. We expect this perturbation to the stress to be caused by the vertical deflection of the membrane whose typical size we denote $\delta$. A simple geometrical argument (see, for example, \cite{Chopin2008}) shows that the induced strain $\propto \delta^2$; since stress is proportional to strain, we therefore expect that $\delta=O(\epsilon^{1/2})$. (Alternatively, the compatibility equation, \cref{Eq:compatibility_dimensionless}, leads to the same scaling.) Hence, we expect that at leading order, the deflection $\zeta=O(\epsilon^{1/2})$. This leading-order amplitude, however, evolves once $t_1=O(1)$, introducing terms at $O(\epsilon^{3/2})$ via the derivative in \cref{eqn:TimeDerivative}. These $O(\epsilon^{3/2})$ terms will introduce a spatial dependence in \cref{Eq:VForceBal_ND} that cannot be balanced by the existing terms (since they are determined at higher order), and so we expect the perturbation to $\zeta(r,t_0,t_1)$ will enter at $O(\epsilon^{3/2})$, rather than $O(\epsilon)$. We therefore let 
\begin{equation}
\zeta(r,t_0,t_1)=\epsilon^{1/2}\zeta_1(r,t_0,t_1)+\epsilon^{3/2}\zeta_3(r,t_0,t_1)+... .
\end{equation}

We must also choose scalings for the damping and the forcing amplitudes, as functions of $\epsilon$. We expect that the damping, $\gamma\partial\zeta/\partial t$, and the forcing, $P$, should enter at the same order of magnitude: the energy inputted via the forcing must be dissipated by the damping. As a result, we expect that $P\sim\epsilon^{1/2}\gamma$.  We also expect that the damping and forcing should enter the problem first through the problem for $\zeta_3$ --- the problem for $\zeta_1$ represents free oscillations of the membrane, and so does not rely on the forcing or damping. As a result, we expect $\gamma=O(\epsilon)$, $P=O(\epsilon^{3/2})$; we therefore let $\gamma=\epsilon\Gamma$ and $P=\epsilon^{3/2}\Pi$.

\subsection{Leading-order problem}

Having chosen these scalings we find that, at leading order:
\begin{equation}
    {\cal L}(\zeta_1)=0,
    \label{eqn:Zeta1WaveEqn}
\end{equation} where the operator ${\cal L}(\cdot)$ is defined by
\begin{equation}
    {\cal L}(\cdot):=\frac{\partial^2(\cdot)}{\partial t_0^2}-\frac{1}{r}\frac{\partial}{\partial r}\left(r\frac{\partial (\cdot)}{\partial r}\right).
\end{equation}

Equation \cref{eqn:Zeta1WaveEqn} describes the free oscillations of the membrane (as expected based on the choice of size of $\gamma$ and $P$) and has solution
\begin{equation}
\zeta_1(r,t_0,t_1)=\left[A(t_1)e^{\ii\omega_0t_0}+\cc\right]J_0(\omega_0r).
\end{equation} Here $\cc$ denotes the complex conjugate,  $J_0(x)$ is the zeroth order Bessel function, and the (dimensionless) fundamental frequency $\omega_0\approx2.4048$ is chosen to ensure that the boundary condition $\zeta(1,t_0,t_1)=0$ is satisfied at leading order.

With this leading order solution for the shape, the leading order problem for the perturbation to the stress is
\begin{equation}
    r\frac{1}{r}\frac{\partial}{\partial r}\left[\frac{1}{r}\frac{\partial(r\psi_2)}{\partial r}\right]=-\frac{\omega_0^2}{2}\left[A(t_1)^2e^{2\ii\omega_0t_0}+|A|^2+\cc\right]\left[J_1(\omega_0r)\right]^2.
\end{equation} We therefore find that
\begin{equation}
\psi_2(r,t_0,t_1)=\left[A(t_1)^2e^{2\ii\omega_0t_0}+|A|^2+\cc\right]\, Y(r)
\label{eqn:psieqn}
\end{equation} 
where $Y(r)$ satisfies 
\begin{equation}
    r\frac{\upd}{\upd r}\left[\frac{1}{r}\frac{\upd}{\upd r}(rY)\right]=-\frac{\omega_0^2}{2}\left[J_1(\omega_0r)\right]^2,
    \label{eqn:CompatGovEq}
\end{equation} subject to boundary conditions
\begin{equation}
    \lim_{r\to0}\bigl[rY'-\nu Y\bigr]=0,\quad Y'(1)=\nu Y(1).
    \label{eqn:CompatBCsG0}
\end{equation} The function $Y(r)$ can be found analytically to be
\begin{equation}
Y(r)=\beta r+\frac{\omega_0^2r}{8} \setlength\arraycolsep{1pt}
{}_1F_2\left(\{1/2\};\{2,2\};-\omega_0^2r^2\right),
\label{eqn:Yr}
\end{equation} where $\setlength\arraycolsep{1pt}
{}_pF_q\left(\{a_1,...,a_p\};\{b_1,...,b_q\};x\right)$ is the generalized hypergeometric function \cite[][]{NISThandbook} and 
\[
\beta=\beta(\nu)=-\frac{\omega_0^2}{8}\setlength\arraycolsep{1pt}
{}_1F_2\left(\{1/2\};\{2,2\};-\omega_0^2\right)+\frac{\omega_0^4}{32(1-\nu)}\setlength\arraycolsep{1pt}
{}_1F_2\left(\{3/2\};\{3,3\};-\omega_0^2\right)
\] is determined from the boundary conditions. Note that the constant $\beta$ depends on the Poisson ratio $\nu$; as a result, the function $Y(r)$ in fact depends on $\nu$ and we therefore denote it $Y(r;\nu)$ henceforth.

\subsection{Next order: The solvability condition}

At $O(\epsilon^{3/2})$, we find that
\begin{align}
{\cal L}(\zeta_3)%\frac{\partial^2\zeta_3}{\partial t_0^2}-\frac{1}{r}\frac{\partial}{\partial r}\left(r\frac{\partial \zeta_3}{\partial r}\right)
=&-2\frac{\partial^2}{\partial t_0\partial t_1}\left[A(t_1)e^{\ii\omega_0t_0}+\cc\right]J_0(\omega_0r)+\frac{\Pi}{2}\left[e^{\ii(\omega_0t_0+\varsigma t_1+\phi)}+\cc\right]\nonumber\\&-\omega_0\Gamma J_0(\omega_0r) \left[\ii A(t_1) e^{\ii\omega_0 t_0}+\cc\right] \nonumber\\
&-\frac{\omega_0}{r}\frac{\partial}{\partial r}\left[Y(r;\nu)J_1(\omega_0r)\right]\left[A^3e^{3\ii\omega_0t_0}+3|A|^2Ae^{\ii\omega_0t_0}+\cc\right].
\label{eq:MS_PDE}
\end{align}

\begin{figure}[!htbp]
    \centering
    \includegraphics[width=0.45\textwidth]{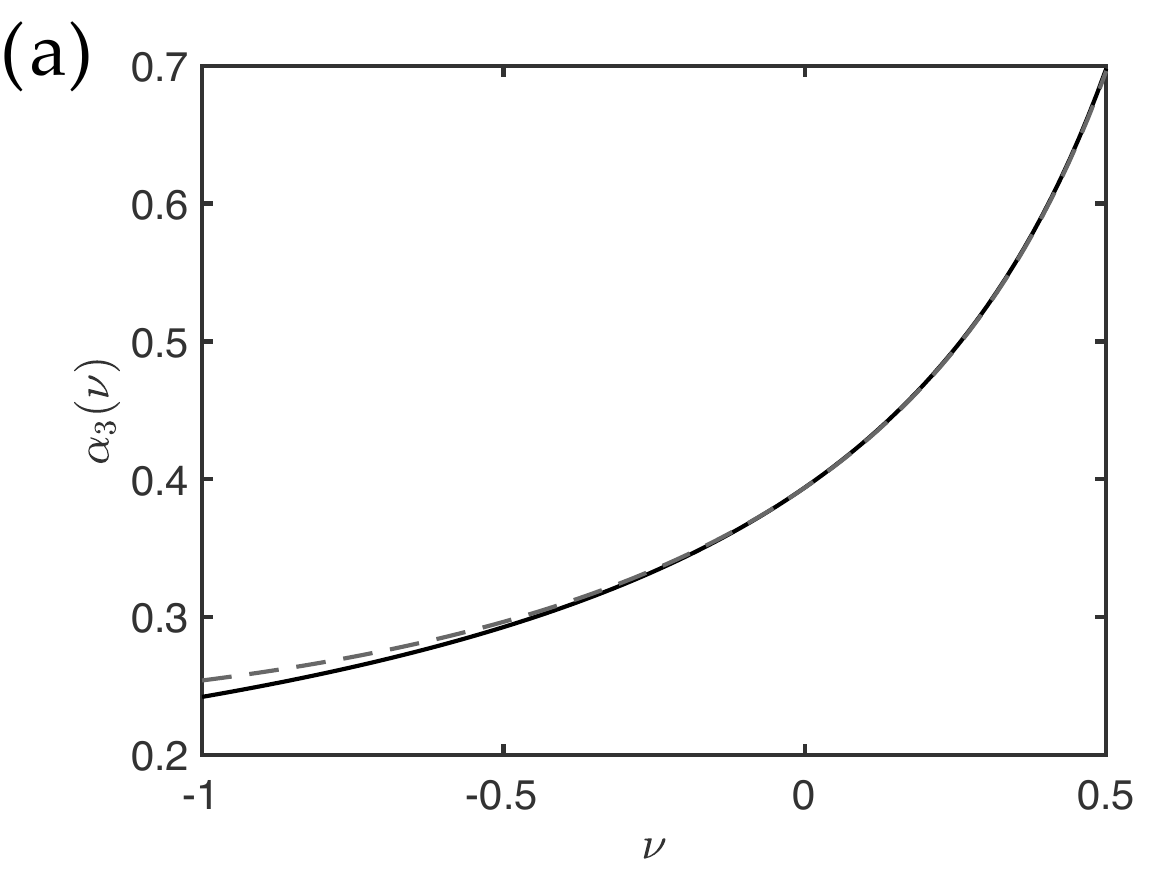}
    \includegraphics[width=0.45\textwidth]{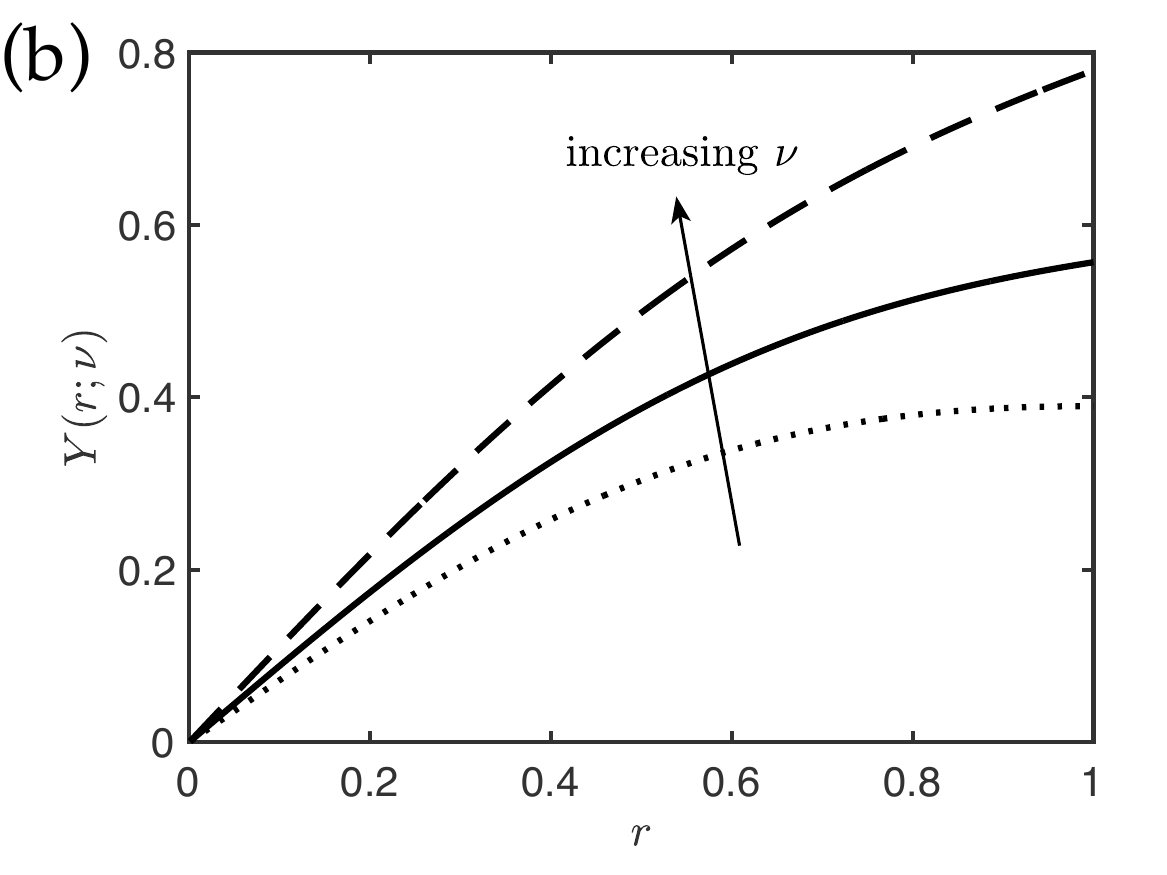}
    \caption{(a) Dependence of the coefficient of the cubic (nonlinear) term, $\alpha_3$, on the Poisson ratio of the membrane, $\nu$,  given exactly by \eqref{eqn:Alpha3exact} (solid curve) compared with the fit to numerical results given by \cite{davidovikj_nonlinear_2017} (dashed curve). (The plot is restricted to the relevant range for isotropic solids, $-1\leq\nu\leq0.5$.)
    (b) The perturbation to the stress function determined  by the method of Multiple Scales, $Y(r;\nu)$ given by \cref{eqn:Yr}, for different values of $\nu$. Results are shown for $\nu=0,0.3$ and $0.5$ with the direction of increasing $\nu$ as indicated.}
    \label{fig:StressPert}
\end{figure}

To make progress, we use the Fredholm Alternative Theorem \cite[][]{Keener}. In the context of a PDE like \cref{eq:MS_PDE}, the Fredholm Alternative Theorem states that either \cref{eq:MS_PDE} has a solution \emph{or} the homogeneous adjoint problem has a solution that is \emph{not} orthogonal to the RHS of \cref{eq:MS_PDE}. The operator ${\cal L}(\cdot)$ on the LHS of \cref{eq:MS_PDE} is self-adjoint with respect to the inner product $\langle u,v\rangle=\int_0^{2\pi/\omega_0}\int_0^1 r\,\overline{u(r,t_0)} v(r,t_0)~\upd r~\upd t_0$. Multiplying the RHS of \cref{eq:MS_PDE} by the complex conjugate of the solution of the homogeneous adjoint problem, 
 $u(r,t_0)=J_0(\omega_0r)e^{\ii\omega_0t_0}$, and integrating, we  find that the condition for a solution of \cref{eq:MS_PDE} to exist (the `solvability condition') is that
\begin{equation}
    0=-2\ii\kone \omega_0\dot{A}+\frac{\Pi\ktwo}{2}e^{\ii(\varsigma t_1+\phi)}-\ii\omega_0\Gamma\kone A - 3\kthree|A|^2A+\cc
    \label{eqn:MS_amplODE}
\end{equation}  where the orthogonality condition gives rise to several constants:
\begin{align}
    \kone&=\int_0^1r\bigl[J_0(\omega_0r)\bigr]^2~\upd r=\frac{[J_1(\omega_0)]^2}{2}\approx 0.1348\nonumber\\
    \ktwo&=\int_0^1rJ_0(\omega_0r)~\upd r=\frac{J_1(\omega_0)}{\omega_0}\approx 0.2159\label{eqn:MS-Ints}\\
    \kthree(\nu)&=\omega_0\int_0^1J_0(\omega_0r)\frac{\upd}{\upd r}\left[Y(r;\nu)J_1(\omega_0r)\right]~\upd r.\nonumber
\end{align} Note that, while the constants $\kone$ and $\ktwo$ are `universal' for this problem, $\kthree$ depends on the Poisson ratio $\nu$. %; in the simulations presented here, we take $\nu=0.3$ and note that $\kthree(0.3)\approx0.5243$. The dependence of $\alpha_3$ on Poisson ratio is shown in fig.~\ref{fig:StressPert}b.
We can calculate this dependence  straightforwardly since the $\nu$-dependence is all encapsulated by the $\beta(\nu)r$ term in $Y(r;\nu)$. We find that
\begin{equation}
\alpha_3(\nu)\approx0.09044+\frac{0.30367}{1-\nu},
\label{eqn:Alpha3exact}
\end{equation} where the approximation is introduced only by taking the constants to five decimal places; the dependence on $\nu$ is exact. In the simulations we present here, $\nu=0.3$ for which $\alpha_3\approx0.524256$. This relationship is shown in fig.~\ref{fig:StressPert}a for completeness.

\paragraph{Comparison to previous results} Note that our analysis gives exact expressions for the constants determined numerically by \cite{davidovikj_nonlinear_2017}. In their notation, we find that $m_{\mathrm{eff}}=2\alpha_1m$, $k_1=2\pi\alpha_1\omega_0^2n_0$, $\xi=2\alpha_2$ and $C_3(\nu)=2\alpha_3(\nu)$, all of which agree with the values given in the Supplementary Information of~\cite{davidovikj_nonlinear_2017}. However, our analytical approach allows for the exact expression given in \eqref{eqn:Alpha3exact} to be determined; the comparison between our analytical expression for $\alpha_3(\nu)$ and the fit proposed by \cite{davidovikj_nonlinear_2017} is shown in fig.~\ref{fig:StressPert}a and shows good agreement over the whole range of $\nu$. However, we emphasize that the technique we have used may, in principle, be generalized to other loadings.

\subsection{Frequency-response curve}

Returning to the solution of \cref{eqn:MS_amplODE}, we write
\begin{equation}
    A(t_1)=\frac{a}{2}e^{\ii\varsigma t_1}
\end{equation} (so that $A+\cc=a\cos\varsigma t_1$ with $a$ the oscillation amplitude).   Taking real and imaginary parts gives equations for the amplitude $a$ and phase shift $\phi$;  $\phi$ may readily be eliminated from these equations giving
\begin{equation}
\frac{\Pi^2\ktwo^2}{a^2}=\kone^2\omega_0^2\Gamma^2+\left(\frac{\kone(\Omega^2-\omega_0^2)}{\epsilon}-\tfrac{3}{4}\alpha_3a^2\right)^2.
    \label{eqn:MS-AmpResponse1}
\end{equation} Recalling  the various scalings with $\epsilon$ (i.e.~the true amplitude $a_1=a\epsilon^{1/2}$, $P=\epsilon^{3/2}\Pi$ and $\gamma=\epsilon\Gamma$) and introducing a rescaled forcing frequency $\bOmega=\Omega/\omega_0$ we immediately have
\begin{equation}
\frac{\ktwo^2}{\kone^2\omega_0^4}\frac{P^2}{a_1^2}=\frac{\gamma^2}{\omega_0^2}+\left(\bOmega^2-1-\tfrac{3}{4}\frac{\kthree}{\kone\omega_0^2}a_1^2\right)^2.
    \label{eqn:MS-AmpResponse2}
\end{equation} 

\Cref{eqn:MS-AmpResponse2} gives a relationship between the oscillation amplitude, $a_1$, and the rescaled forcing frequency $\bOmega$; note that the form of \cref{eqn:MS-AmpResponse2} has precisely the same {form as the amplitude response for the Duffing equation derived using Multiple Scales \cite[][]{nayfeh_nonlinear_2008,LifshitzReview}, which has been used to fit various experimental parameters previously (as discussed in the Introduction).} However, unlike an informal analysis based on the Duffing oscillator, the constants have been derived formally via an asymptotic analysis of the governing PDEs that is valid in the limit of small oscillation amplitudes. This approach means that, in principle, no fitting of the constants should be required to give excellent agreement between the numerically-determined response curves and the predictions of \cref{eqn:MS-AmpResponse2} in the limit of small amplitude oscillations.

We note that \eqref{eqn:MS-AmpResponse2} is a quadratic equation in $\bOmega^2$; it is therefore a straightforward exercise to determine $\bOmega^2$ as a function of $a_1$ and $P$ --- i.e.~to determine the frequency required to give oscillations of a particular amplitude. This analysis shows that a branch point in $\bOmega$ occurs when
\begin{equation}
    \frac{a_1}{P}=\frac{\ktwo}{\gamma\omega_0\kone}.
    \label{eqn:MaxComplianceMS}
\end{equation} Note then that this `maximum compliance', $(a_1/P)_{\max}$, is independent of the amplitude of the forcing $P$.

\subsection{Stress perturbation}

An advantage of our approach is that it gives more detailed information about the effects of deformation; for example, the above calculation allows us to determine the  perturbation to the stress caused by vibration: the function $Y(r;\nu)$ is defined in \cref{eqn:psieqn} in terms of the perturbation to the stress function $\psi_2(r,t)=\bigl[\psi(r,t)-r\bigr]$. The analytical prediction of $Y(r;\nu)$, given by  \cref{eqn:Yr}, is shown for three different values of $\nu$  in fig.~\ref{fig:StressPert}b. 

Having presented the predictions of a Multiple Scales analysis, we now turn to a numerical resolution of the full system, namely \cref{Eq:VForceBal_ND}--\cref{eqn:PressureImposed}.

\section{Numerical method and results\label{sec:Numerics}}

We determine the dynamic evolution of the membrane by solving \cref{Eq:VForceBal_ND,Eq:compatibility_dimensionless,Eq:bcs_dimensionless} numerically, subject to the forcing in \cref{eqn:PressureImposed} with phase $\phi=0$ chosen for simplicity. 
The PDE, \cref{Eq:VForceBal_ND} is numerically solved by using the method of lines with numerical integration performed in Python using the \texttt{solve\_ivp} routine to advance the shape of the membrane in time. At each time step, the stress state must also be updated in response to the shape; this is a boundary value problem and hence is solved using \texttt{solve\_bvp}. Details of this procedure are presented in \ref{sec:AppendixNums}. 

\begin{figure}[!htbp]
    \centering
    \includegraphics[width=0.9\textwidth]{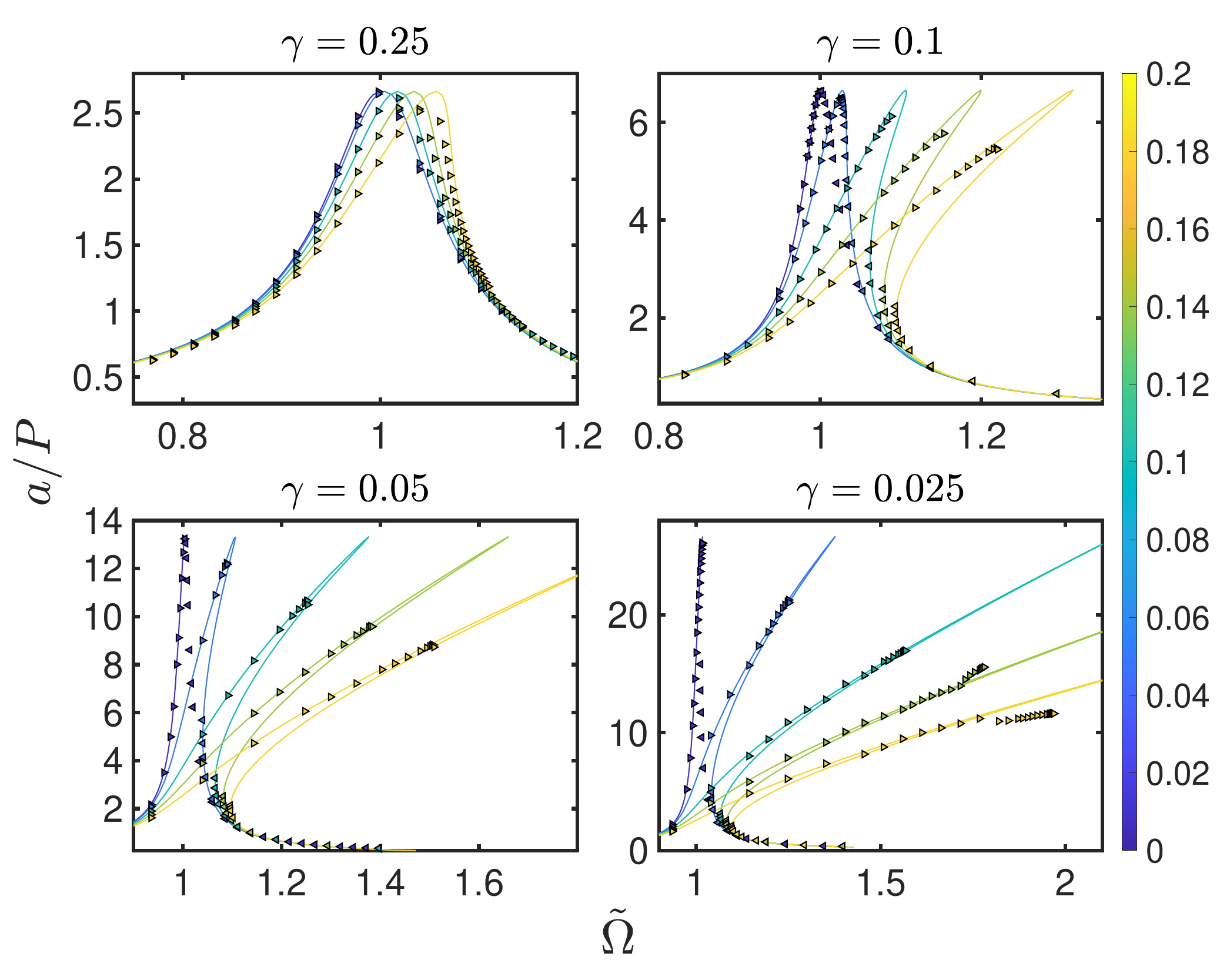}
    \caption{Detailed response curves with a range of values of dimensionless damping, $\gamma$. The solid curves show the response predicted by the method of Multiple Scales, \cref{eqn:MS-AmpResponse2}, without any fitting. In each plot,  triangular points represent numerical results obtained at different driving pressures (as indicated by the colourbar to the right). The right-pointing triangles indicate a forward sweep, while the left-pointing triangles indicate a backward sweep of frequency. }
    \label{fig:DetailedResponseMS}
\end{figure}

The numerical results show that, after an initial transient, the vibrations of the membrane reach an approximately periodic state, with some  amplitude $a$. These results also show that the value of $a$ is affected by both the applied pressure and the applied frequency. This is precisely the information captured by a frequency--response curve. We generate the frequency--response curve by a forward sweep, i.e.~we increase the frequency by small increments starting from a frequency close to zero.  The first and leftmost point on the curve is obtained with this small forcing frequency, assuming that the membrane starts from rest. Then the frequency is increased by  a small step. The initial conditions of the current step  is the membrane displacement and velocity corresponding to the instant of time at which the frequency was increased from the previous step \footnote{We have found that, when increasing the forcing frequency, it is important to maintain the phase of the driving oscillation, rather than starting back from $t=0$; otherwise the discontinuity introduced by the change of phase can lead to premature `jump-down'.}.  As we continue the forward sweep, the amplitude reaches a maximum value $a_\mathrm{max}$: a further increase in frequency ($\tilde{\Omega}>\tilde{\Omega}_\mathrm{max}$)  causes the amplitude to suddenly jump down to a significantly smaller value. Therefore, we must ensure that the increment used for the sweep is sufficiently small to capture the correct maximum amplitude. The markers in fig.~\ref{fig:DetailedResponseMS} correspond to adaptive forward sweeps with variable frequency increments as the sweep progresses, with each colour representing a different magnitude of pressure. Our sweep is adaptive in the sense that, when a jump-down occurs, we revisit the previous step, sweeping forward with a smaller frequency increment. We do this as long as the increment size is larger than a specified tolerance  ($\Delta \bOmega>10^{-6}$). For completeness, we also perform an adaptive reverse sweep starting with values of $\bOmega>\bOmega_{\max}$, decreasing $\bOmega$ until the system jumps up.

As might be expected, jump-down happens at larger amplitudes for larger pressure. To normalize for this effect, fig.~\ref{fig:DetailedResponseMS} shows the compliance, $a/P$, plotted as a function of $\bOmega$. Recall from the discussion around \eqref{eqn:MaxComplianceMS} that the results of the Multiple Scales analysis suggests that the maximum compliance should be independent of the loading $P$. However, our numerical results show that  the maximum compliance varies with forcing. This  is often associated with nonlinear damping \cite[][]{LifshitzReview}, even though  our numerical simulations use only linear damping.

\begin{figure}[!htbp]
    \centering
    \includegraphics[width=0.45\textwidth]{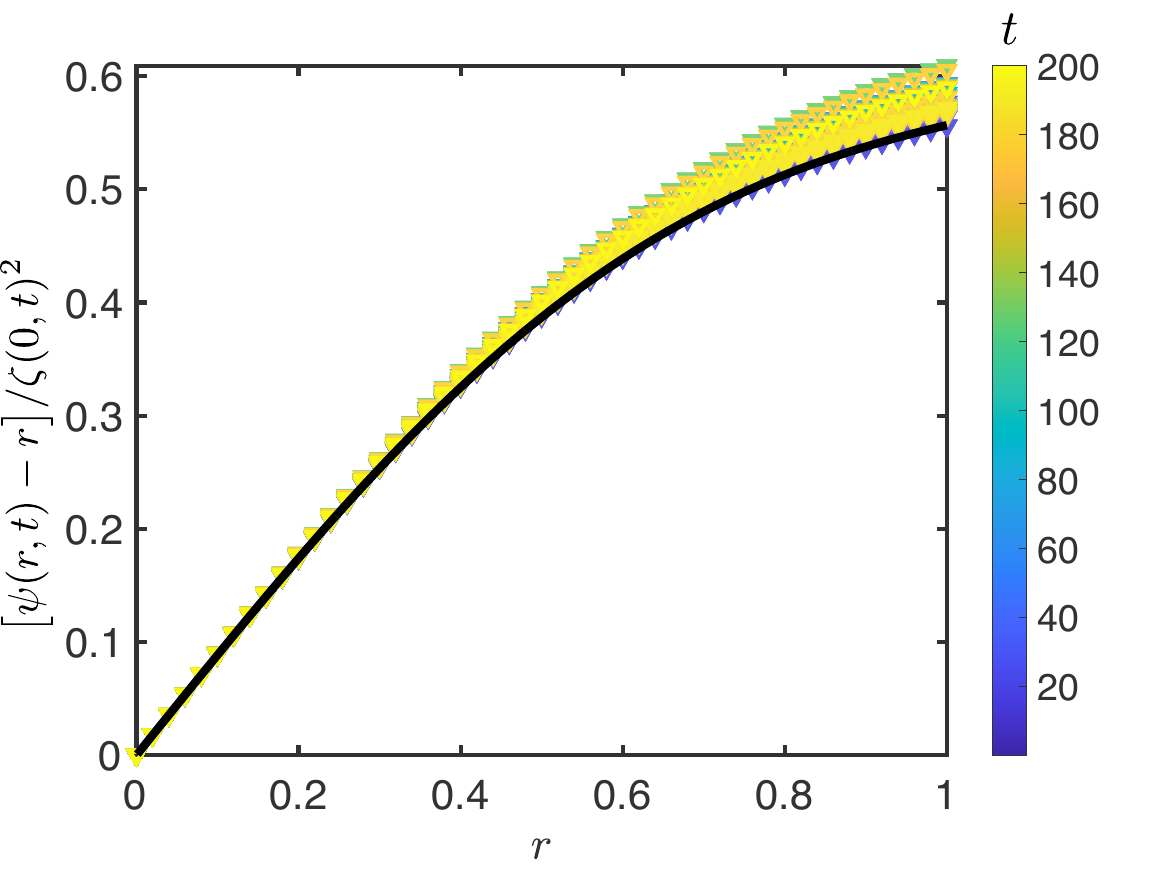}
    \caption{The perturbation to the stress function determined numerically (coloured points) and predicted by the method of Multiple Scales, $Y(r;\nu)$ given by \cref{eqn:Yr}, (solid curve). Here snap-shots of this perturbation are shown at various instants of time (indicated by point colour, see colourbar to the right) while the constants $\gamma=0.1$, $P=0.1$ and $\bOmega=1.04$.}
    \label{fig:StressPertNums}
\end{figure}

As another test of our numerical procedure and the Multiple Scales analysis, fig.~\ref{fig:StressPertNums} shows a comparison between the analytical prediction for the perturbation to the stress function, $Y(r;\nu)$ given by  \cref{eqn:Yr},  and an estimate of this function obtained from our numerical results. This shows reasonable, if not perfect, agreement between the two (independent) approaches.

Our comparison between the numerical results and the predictions of our Multiple Scales analysis has shown that the results are generally in good agreement for small amplitude (either with low forcing, or large damping). However, the jump-down behaviour predicted by the Multiple Scales method is far from what is observed numerically. We now turn to consider how the predictions of an alternative approach, which uses a Galerkin approach together with Harmonic Balance, can give a better account of our numerical results.

\section{Alternative theory: Galerkin approach and Harmonic Balance}\label{sec:HB}

\subsection{Galerkin representation}

As an alternative point of view, we consider a modal decomposition of the problem --- a Galerkin approach --- in which we expand the membrane deformation as a series in terms of the normal modes of the membrane, i.e.~we let
\begin{equation}
    \zeta(r,t)=\sum_{n=0}^\infty A_n(t)J_0(\omega_nr)
    \label{eqn:GalerkinZeta}
\end{equation} in which  the $A_n(t)$ are unknown amplitudes and the $\omega_n$ are roots of $J_0(\omega_n)=0$; in particular, $\omega_0\approx2.40483$, $\omega_1\approx 5.52008$ and $\omega_2\approx8.65373$ are the dimensionless frequencies of the first three modes.

Of course, in addition to the decomposition \cref{eqn:GalerkinZeta}, we must choose an appropriate decomposition of the stress function $\psi(r,t)$. In this regard, we note that upon substituting \cref{eqn:GalerkinZeta} into the dimensionless compatability equation \cref{Eq:compatibility_dimensionless} we have
\begin{equation}
    r\frac{\partial}{\partial r}\left[\frac{1}{r}\frac{\partial(r\psi)}{\partial r}\right]=-\frac{1}{2}\left[\sum_{n=0}^\infty \omega_nA_n(t)J_1(\omega_nr)\right]^2.
\end{equation} This form motivates introducing the expansion
\begin{equation}
    \psi(r,t)=r+\sum_{m,n=0}^\infty A_m(t)A_n(t)Y_{m,n}(r)
    \label{eqn:GalerkinPsi}
\end{equation}
where the functions $Y_{m,n}(r)$ satisfy
\begin{equation}
    r\frac{\upd}{\upd r}\left[\frac{1}{r}\frac{\upd}{\upd r}(rY_{m,n})\right]=-\begin{cases}
        \frac{\omega_m^2}{2}\left[J_1(\omega_mr)\right]^2,& m=n,\\
        \omega_m\omega_n\,J_1(\omega_mr)J_1(\omega_nr), &m\neq n,
    \end{cases}
    \label{eqn:YmnEqn}
\end{equation} subject to boundary conditions
\begin{equation}
    \lim_{r\to0}\bigl[rY_{m,n}'-\nu Y_{m,n}\bigr]=0,\quad Y_{m,n}'(1)=\nu Y_{m,n}(1).
    \label{eqn:YmnBCs}
\end{equation} (Note that we are not able to give explicit formulae for $Y_{m,n}$ when $m\neq n$; when $m=n$ the form from \cref{eqn:Yr} holds with $\omega_0\to \omega_m$.)

Substituting the expansions \cref{eqn:GalerkinZeta} and \cref{eqn:GalerkinPsi} into the vertical force balance equation \cref{Eq:VForceBal_ND} we find that
\begin{equation}
    \sum_{n=0}^\infty\left(\ddot{A}_n+\gamma\dot{A}_n+\omega_n^2A_n\right)J_0(\omega_nr)=p(r,t)-\frac{1}{r}\frac{\partial}{\partial r}\left(\sum_{l,m,n}\omega_lA_lA_mA_nJ_1(\omega_lr)Y_{m,n}(r)\right).
    \label{eqn:GalerkinVFB}
\end{equation}

At this stage, it is natural to simplify the LHS of \cref{eqn:GalerkinVFB} by projecting using the operator $\int_0^1rJ_0(\omega_m r)(\cdot)~\upd r$ and using the orthogonality relationship $\int_0^1rJ_0(\omega_m r)J_0(\omega_nr)~\upd r=\delta_{mn}[J_1(\omega_m)]^2/2$. We therefore find that
\begin{equation}
    \ddot{A}_k+\gamma\dot{A}_k+\omega_k^2A_k=\frac{1}{\alpha_{1,k}}\left\{\alpha_{2,k}p(t)-\sum_{l,m,n}\alpha_{3,klmn}A_lA_mA_n\right\}
    \label{eqn:GalerkinCoupled}
\end{equation} where we have assumed that $p(r,t)=p(t)$ is spatially uniform and
\begin{eqnarray}
\alpha_{1,k}&=\int_0^1r[J_0(\omega_kr)]^2~\upd r=\frac{[J_1(\omega_k)]^2}{2},\label{eqn:alpha1k}\\
    \alpha_{2,k}&=\int_0^1rJ_0(\omega_kr)~\upd r=\frac{J_1(\omega_k)}{\omega_k},\\
    \alpha_{3,klmn}&=\omega_l\int_0^1J_0(\omega_kr)\frac{\upd }{\upd r}\left[J_1(\omega_lr)Y_{m,n}(r)\right]~\upd r.\label{eqn:alpha3klmn}
\end{eqnarray} It is worth noting that the expressions for the constants given in \eqref{eqn:alpha1k}--\eqref{eqn:alpha3klmn} are generalizations of the constants determined in \eqref{eqn:MS-Ints} for the fundamental frequency, $\omega_0$.

\begin{figure}
    \centering
    \includegraphics[width=0.5\linewidth]{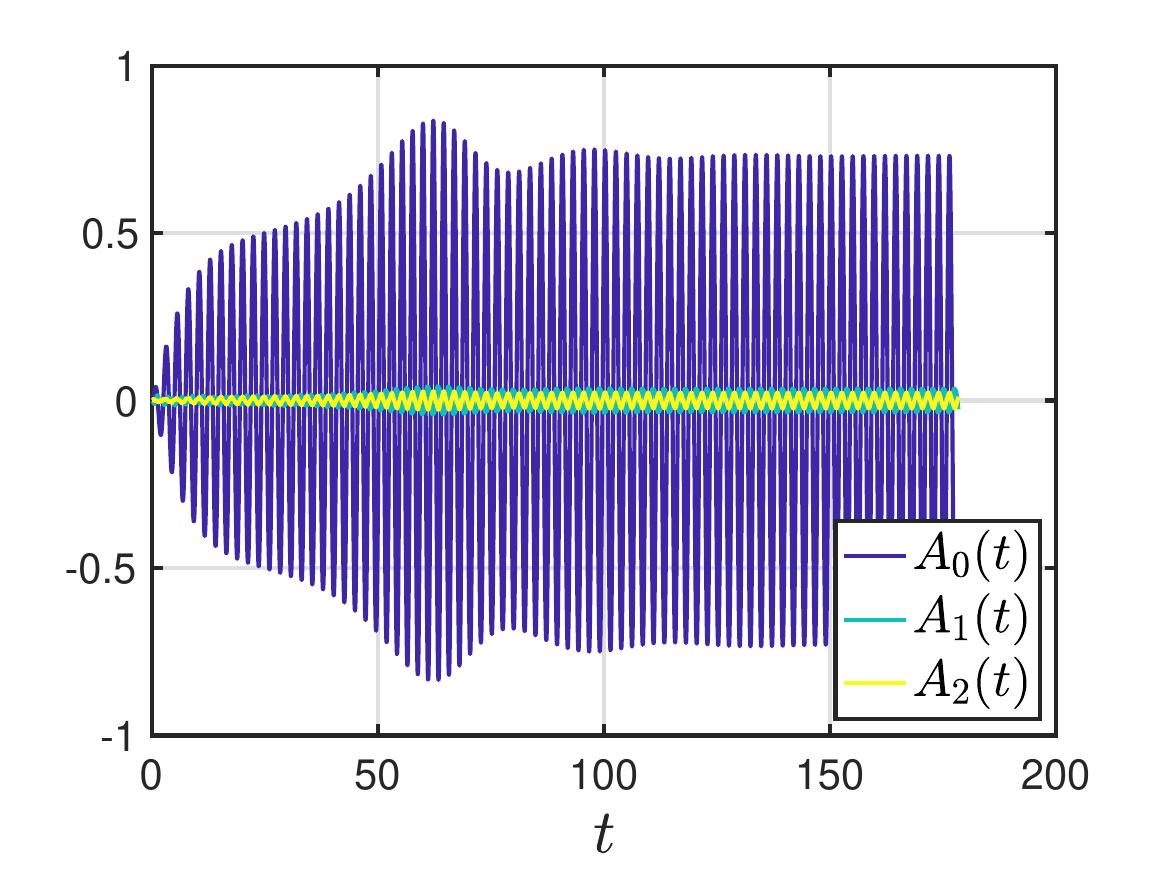}
    \caption{Evolution of the first three modes for $\gamma=0.1$, $P=0.18$ and $\bOmega\approx 1.098$. We see that the fundamental mode, with dimensionless frequency $\omega_0$, is clearly dominant since $A_0\gg A_1,A_2$. This justifies our neglect of the mode-coupling in this problem.}
    \label{fig:CompareModeAmplitudes}
\end{figure}

To make further progress, we shall consider henceforth, only the $k=0$ mode and, further will assume that the only relevant nonlinear term comes from the $A_0^3$ term. This will simplify the resulting analysis considerably.
To motivate this simplification, we calculate the first three amplitudes from our full numerical simulation of the problem for parameter values that are well into the nonlinear (deformation-induced strain) regime, see fig.~\ref{fig:CompareModeAmplitudes}. (These amplitudes are calculated from the numerically determined $\zeta(r,t)$ via the integrals:
\begin{equation}
    A_n(t)=\frac{\int_0^1r\zeta(r,t)J_0(\omega_n r)~\upd r}{\int_0^1r[J_0(\omega_n r)]^2~\upd r},
\end{equation} for $n=0,1,2$.) We  see from this plot that only the lowest mode, dimensionless frequency $\omega_0$, has a significant non-zero amplitude during the oscillation, even though the parameters in this case keep us well within the nonlinear regime. To proceed we therefore simplify \cref{eqn:GalerkinCoupled} by neglecting the coupling between modes, and only consider the dominant fundamental mode, with $k=0$; this  gives
\begin{equation}
    \ddot{A}_0+\gamma\dot{A}_0+\omega_0^2A_0+\frac{\alpha_{3}}{\alpha_1}A_0^3=\frac{\alpha_{2}}{\alpha_{1}}p(t).
    \label{eqn:GalerkinDecoupled}
\end{equation} We now turn to study this equation using the method of Harmonic Balance.

\subsection{Harmonic Balance}

\begin{figure}[!htbp]
    \centering
    \includegraphics[width=0.9\textwidth]{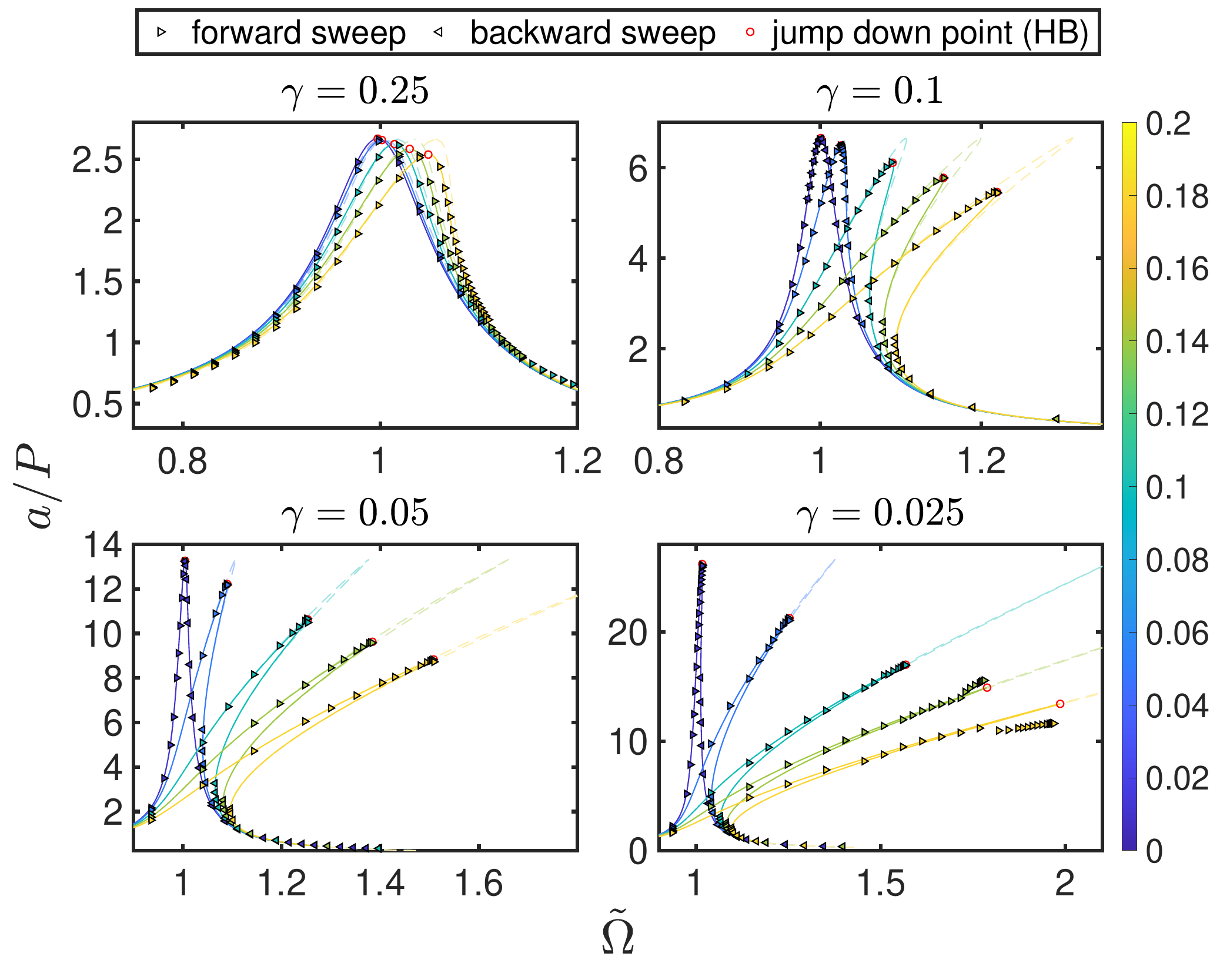}
    \caption{Detailed response curves with a range of values of dimensionless damping, $\gamma$. The solid curves represent the predictions of the Harmonic Balance method with no fitting parameters. This shows better agreement with the numerical results than did the method of Multiple Scales (fig.~\ref{fig:DetailedResponseMS}, but also light dashed curves here). In each plot, the red circles indicate the jump-down point predicted by Harmonic Balance, triangular points represent numerical results obtained at different driving pressures (as indicated by the colourbar to the right). The right-pointing triangles indicate a forward sweep, while the left-pointing triangles indicate a backward sweep of frequency. }
    \label{fig:DetailedResponseHB}
\end{figure}

In Harmonic Balance, the main idea is to pose a solution of the form $A_0(t)=a_0\cos\Omega t$ and to determine the amplitude $a_0$ from requiring the $\cos\Omega t$ behaviour to match with the forcing.  With the forcing from \cref{eqn:PressureImposed} (which incorporates a phase difference $\phi$ between forcing and response), we apply the method of Harmonic Balance to \cref{eqn:GalerkinDecoupled} and eliminate the phase $\phi$, leading to  
\begin{equation}
\frac{\ktwo^2}{\kone^2\omega_0^4}\frac{P^2}{a_0^2}=\frac{\gamma^2}{\omega_0^2}\bOmega^2+\left(\bOmega^2-1-\tfrac{3}{4}\frac{\kthree}{\kone\omega_0^2}a_0^2\right)^2.
    \label{eqn:HB_AmpResponse}
\end{equation} This result is used in the plots of fig.~\ref{fig:DetailedResponseHB} and shows very good agreement with the numerically determined frequency--amplitude response curves. 

\subsection{Multiple Scales \emph{versus} Harmonic Balance}

The result of the Galerkin approximation and Harmonic Balance, \cref{eqn:HB_AmpResponse}, is very similar to the corresponding result from the Multiple Scales analysis, \cref{eqn:MS-AmpResponse2}. Indeed, the only difference is the first term on the RHS, which represents the viscous damping and is independent of the driving frequency $\bOmega$ according to the Multiple Scales analysis, but depends on $\bOmega$ according to the Harmonic Balance analysis. This difference is generally small  (since $\bOmega\approx1$ after all) but becomes particularly important close to the point of maximum compliance and the jump-down point. In particular, the prediction that the maximum compliance should be independent of load strength \cite[see][for example]{LifshitzReview} is only true of the Multiple Scales approach; as a result, non-constant maximum compliance does not necessarily indicate the presence of nonlinear damping, but could instead be explained as a breakdown in the applicability of the Multiple Scales analysis.

The Harmonic Balance method is expected to more accurately describe the solutions of strongly nonlinear differential equations  than perturbation methods \cite[][]{Lim2003, Lai2009}. It is perhaps then not surprising that \cref{eqn:HB_AmpResponse} shows a better fit to the amplitude response in our numerical simulations: the decoupled equation \cref{eqn:GalerkinDecoupled} is a Duffing-type equation, for which the Harmonic Balance method is known to perform well, even with a single harmonic \cite[][]{Krack.2019}.

\begin{figure}
    \centering
    \includegraphics[width=0.8\linewidth]{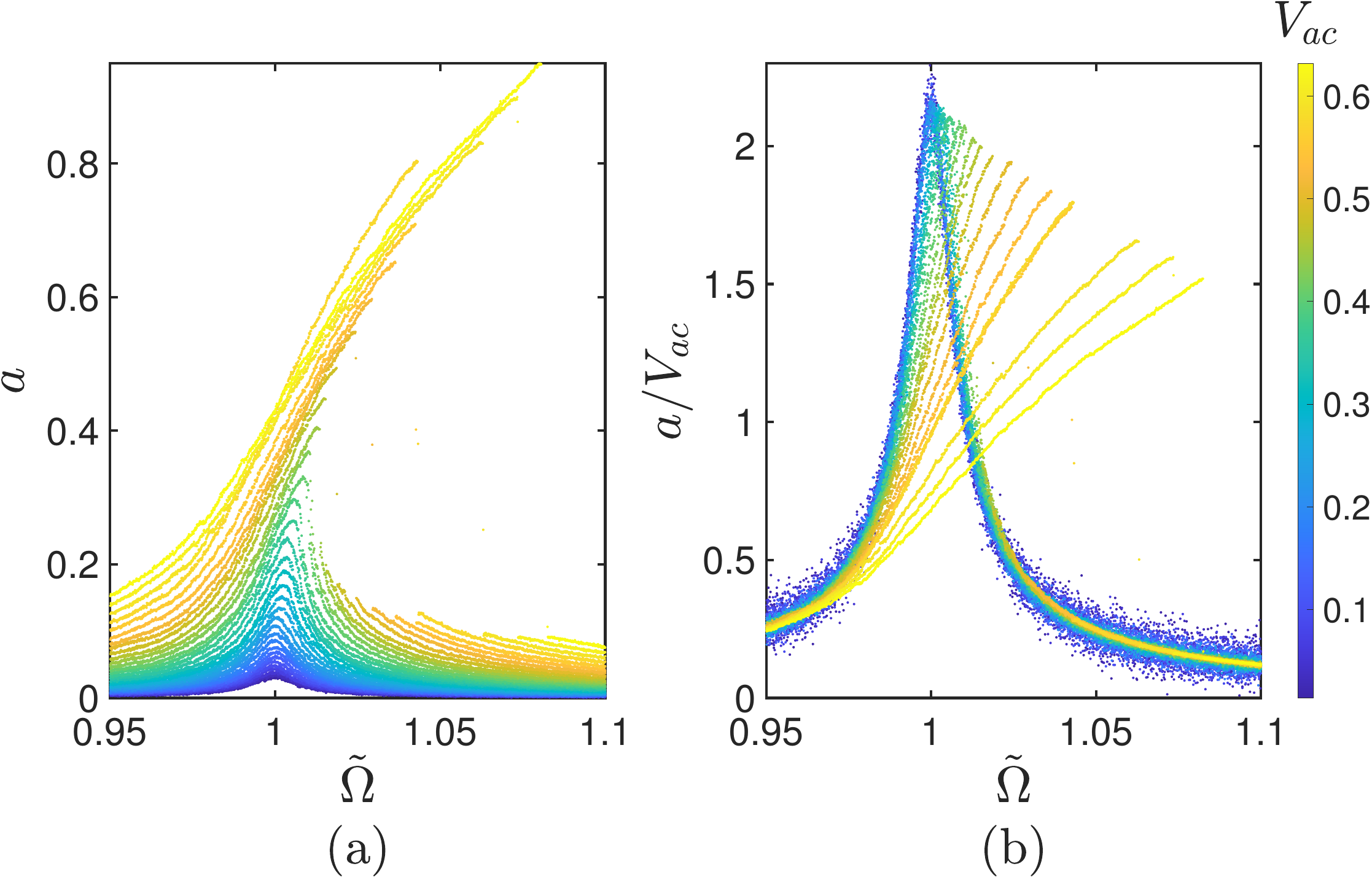}
    \caption{Experimental response curves for  different driving voltages, $V_{ac}$, obtained by \cite{davidovikj_nonlinear_2017}. (a) Raw response curves show the oscillation amplitude $a$, normalized by sheet thickness, versus rescaled driving frequency $\bOmega=\Omega/\Omega_0$ for different applied voltages $V_{ac}$. (b) The compliance/responsivity of the results shows that the maximum compliance decreases as the driving voltage increases. (The magnitude of the driving voltage is shown by point colour, as shown in the colourbar to the right.)}
    \label{fig:ExptData}
\end{figure}

Having discussed the effect, and origin, of differences between the two analytical approaches discussed here to derive reduced models of the oscillations of a vibrating membrane subject to a uniform pressure, it is natural to wonder how the predictive power of the  two models can be compared to experiments. Given  experimental data, such as that of \cite{davidovikj_nonlinear_2017} for example, how can we assess whether discrepancies between the observed and expected force--amplitude response are caused by poor fitting or by additional physics in the experiments, such as nonlinear damping? In terms of distinguishing the suitability of the two approaches discussed here, we note that experimental data frequently exhibits a non-constant maximum compliance. For example, data from \cite{davidovikj_nonlinear_2017} shows a decreasing maximum compliance when the oscillation amplitude is normalized by the driving AC voltage, see fig.~\ref{fig:ExptData}. As a result,  \cref{eqn:HB_AmpResponse} is more likely to be an appropriate frequency-response relation than \cref{eqn:MS-AmpResponse2}.  The question then becomes how to do the fitting of the unknown parameters in \cref{eqn:HB_AmpResponse}, particularly $\gamma$. This is usually done via the jump-down point, and so we now discuss some considerations for this procedure that are highlighted by our analysis.

\section{Advice for fitting\label{sec:fitting}}

\begin{figure}[!htbp]
     \centering
     \begin{subfigure}[b]{0.45\textwidth}
         \centering
         \includegraphics[width=\textwidth]{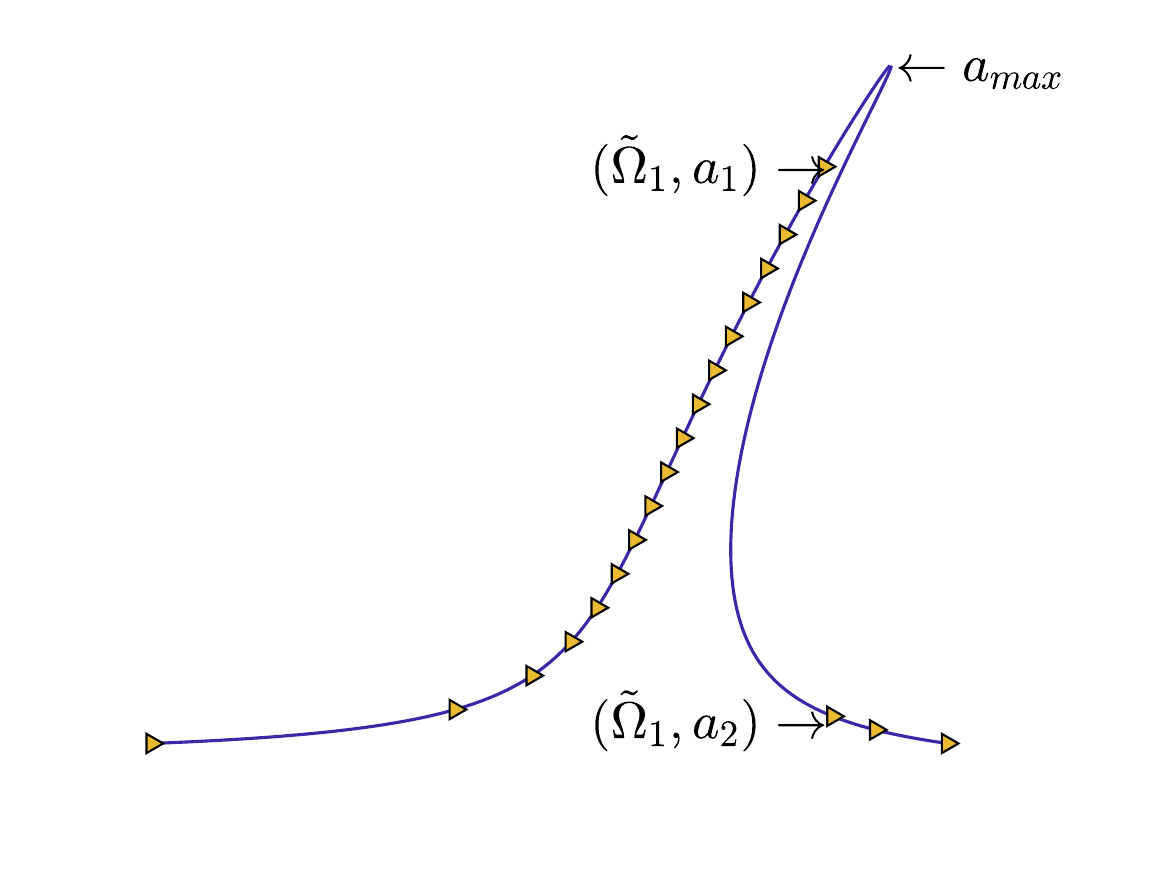}
         \caption{}
     \end{subfigure}
     \hfill
     \begin{subfigure}[b]{0.45\textwidth}
         \centering
         \includegraphics[width=\textwidth]{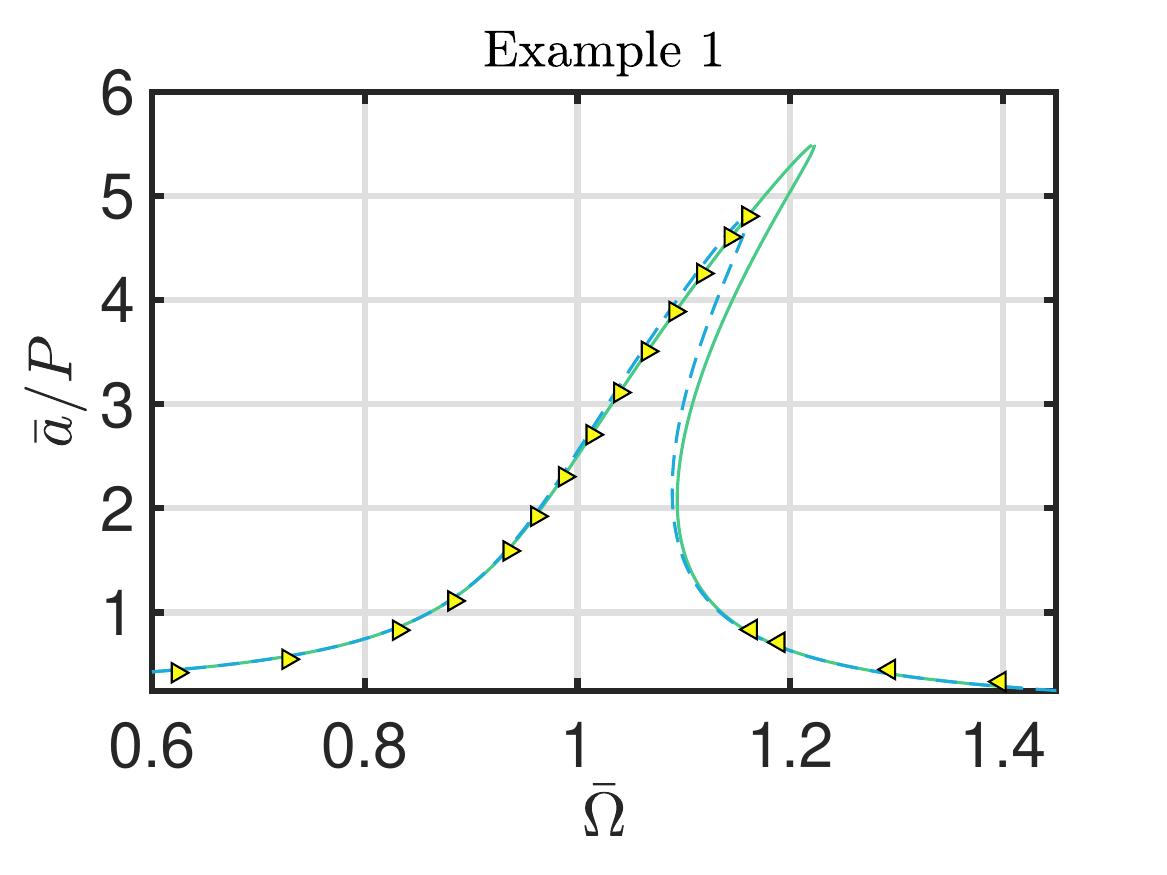}
         \caption{}
     \end{subfigure}
        \caption{(a) Schematic of experimental data (yellow triangles) overlaid on analytical amplitude response (solid curve) with a (premature) jump-down from $(\bOmega_1,a_1)$ to $(\bOmega_1,a_2)$ simulated by truncating the data set. (b) An illustrative example to compare the errors induced by fitting a data set with early jump-down. Here, jump down is observed at $a_1=0.8647$, $a_2=0.1475$, $\bOmega_1=1.16$; the plot is generated by truncating data from our numerical simulations with $\gamma=0.1$, $\nu=0.3$ and $P=0.18$. The values of $\gamma$ and $\alpha_3(\nu)$ are then fitted by two procedures: (i) using the two points at the jump-down frequency $\bOmega_1$ leads to the frequency--response curve as a solid curve while (ii) assuming the jump-down point corresponds to the maximum point of the frequency--response curve leads to the dashed curve, which is a significantly worse fit of the data over most of the curve.}
    \label{fig:prematurejumpdown}
\end{figure}

In prior investigations within this domain, experimental data is commonly fitted to the theoretically-determined frequency--response data by matching the maximum amplitude (i.e.~the point at which `jump-down' occurs). However, ample literature attests to the potential for jump-down to happen prematurely (see schematic in fig.~\ref{fig:prematurejumpdown}a) during experiments, whether because of noise in the system or other practical constraints.    For example, \cite{Chowdhury2017,Chowdhury2020} report that  when the system is close to the peak amplitude of  the response curve, the system is sensitive to inherent noise in the experimental system, and suffers an unexpected early jump-down. We also observe early jump-down in other published experimental data  \cite[see][for example]{Mestrom2008,Weber2014} and surmise that this may be due to similar effects.

 Figure \ref{fig:prematurejumpdown}a shows a schematic of premature jump-down, which occurs at $(\bOmega_1,a_1)$, instead of the point of maximum amplitude, $(\bOmega_{\max},\amax)$. One might therefore expect a fit based on the assumption that $a_1$ is the maximum amplitude to lead to a significant fitting error. Instead, we note that at this jump-down point  the amplitude drops to $a_2$ from $a_1$ with only a negligible change in the forcing frequency. Based on the results presented in Section \ref{sec:theory}, particularly \eqref{eqn:HB_AmpResponse}, we suggest that the frequency at the jump-down point, regardless of whether it happens prematurely or not, together with the amplitude it jumps to, contains enough information for a more  robust way to fit experimental data to the analytical response curves. We can do this by treating \eqref{eqn:HB_AmpResponse} as a quadratic in $\bOmega^2$ and solving;
  %\cref{Eq:bAndCFromHB}
 using the  two amplitudes at the jump-down frequency, $\bOmega_1$ we find that $\bOmega$, $a_1$, $a_2$, $\gamma$ and $\alpha_3$ satisfy
\begin{align}
    \bOmega_{1}^2&=1+\tfrac{3}{4}\frac{\kthree}{\kone\omega_0^2}a_1^2-\frac{\gamma^2}{2\omega_0^2}-\sqrt{\left(1+\tfrac{3}{4}\frac{\kthree}{\kone\omega_0^2}a_1^2-\frac{\gamma^2}{2\omega_0^2}\right)^2-\left(1+\tfrac{3}{4}\frac{\kthree}{\kone\omega_0^2}a_1^2\right)^2+\frac{\ktwo^2}{\kone^2\omega_0^4}\frac{P^2}{a_1^2}} \label{Eq:correctFita1}\\
    \bOmega_{1}^2&=1+\tfrac{3}{4}\frac{\kthree}{\kone\omega_0^2}a_2^2-\frac{\gamma^2}{2\omega_0^2}+\sqrt{\left(1+\tfrac{3}{4}\frac{\kthree}{\kone\omega_0^2}a_2^2-\frac{\gamma^2}{2\omega_0^2}\right)^2-\left(1+\tfrac{3}{4}\frac{\kthree}{\kone\omega_0^2}a_2^2\right)^2+\frac{\ktwo^2}{\kone^2\omega_0^4}\frac{P^2}{a_2^2}}\label{Eq:correctFita2}
\end{align}
Fitting \cref{Eq:correctFita1,Eq:correctFita2} to any experimental data involves finding the unknowns $\gamma$ and $\alpha_3$, and can be solved numerically. Please note that \cref{Eq:correctFita1,Eq:correctFita2} are dimensionless and if we were to consider its dimensional form, the fitting is the equivalent of obtaining $\bar{c}$ and $\alpha_3 E_{2D}$ from \cref{Eq:correctFita1D,Eq:correctFita2D} below. 

For simplicity, we focus on \cref{Eq:correctFita1,Eq:correctFita2} and compare the error introduced in calculations by assuming that the jump-down point corresponds to the maximum amplitude possible. At the maximum amplitude, $\bOmega_1$, $a_1$, $a_2$, $\gamma$ and $\alpha_3$ should satisfy
\begin{align}
a_{1}^2&=\frac{2\kone \omega_0^2}{3\kthree}\left\{\frac{\gamma^2}{4\omega_0^2}-1+\left[\left(1-\frac{\gamma^2}{4\omega_0^2}\right)^2+\frac{3\ktwo^2\kthree P^2}{\kone^3\gamma^2\omega_0^4}\right]^{1/2}\right\} \label{Eq:incorrectFita1}\\
\bOmega_{1}^2&=1+\tfrac{3}{4}\frac{\kthree}{\kone\omega_0^2}a_2^2-\frac{\gamma^2}{2\omega_0^2}+\sqrt{\left(1+\tfrac{3}{4}\frac{\kthree}{\kone\omega_0^2}a_2^2-\frac{\gamma^2}{2\omega_0^2}\right)^2-\left(1+\tfrac{3}{4}\frac{\kthree}{\kone\omega_0^2}a_2^2\right)^2+\frac{\ktwo^2}{\kone^2\omega_0^4}\frac{P^2}{a_2^2}} \label{Eq:incorrectFita2},
\end{align}
instead of \cref{Eq:correctFita1,Eq:correctFita2}.

To illustrate  the difference in two approaches, consider the `experimental' data denoted by yellow triangles in \cref{fig:prematurejumpdown}b. These data were generated numerically as before but {truncated prior to} the point of jump-down to mimic experimental data with premature jump-down.%\st{by performing numerical experiments with a different sweeping protocol (?)} 
 The data was generated with  known values of the parameters; in particular, $P=0.18$, $\gamma=0.1$ and $\nu=0.3$ (so that $\alpha_3\approx0.524256$).  For this data we observe $a_1\approx0.8647$, $a_2\approx0.1475$, and $\bOmega_1\approx1.16$ at the simulated jump-down point (which we do \emph{not} assume corresponds to the maximum amplitude possible). In principle, the values of $\gamma$ and $\alpha_3$ are not known (for example if the Poisson ratio $\nu$ is unknown), but all other parameters are either known or are measured directly from the frequency--response curve. We fit the data to \cref{Eq:correctFita1,Eq:correctFita2} by varying $\gamma$ and $\alpha_3$ and using \texttt{fsolve} in MATLAB; we find $\gamma\approx0.0987$ and $\alpha_3\approx0.5255$. Note that the fitted values of $\gamma$ and $\alpha_3$ are both within 2\% of the true values in the simulations. The corresponding frequency--response curve is shown by the solid green curve in \cref{fig:prematurejumpdown}b, and matches well the numerical data over the whole frequency spectrum with a  residual sum of squares of $0.0079$. By contrast, if we had assumed that jump-down happens at the point of maximum amplitude and used \cref{Eq:incorrectFita1,Eq:incorrectFita2}, we would have found  $\gamma\approx0.1196$ and   $\alpha_3\approx0.4785$. These values differ from the true values by 10-20\%. Moreover, the corresponding frequency--response curve, shown as  the dashed blue curve in \cref{fig:prematurejumpdown}b, deviates from the experimental data points even away from the maximum --- the residual sum of squares is now 0.0425, significantly higher than for the frequency--response curve that was fitted based on the two pieces of information from the observed jump-down point. 

This refinement to the fitting procedure is important because an error in estimating  $\alpha_3$ would then lead to error in the calculation of material properties $E_{2D}$ and $\nu$, which is the motivating factor for calculations. This becomes  clearer when we examine \cref{Eq:correctFita1,Eq:correctFita2} in their dimensional form, which can be obtained by using \cref{Eq:NDvariables,Eq:normal_p,Eq:gamma} and by substituting $\bOmega = \bar{\Omega}/\left(\omega _0 \sqrt{\frac{T_{\text{pre}}}{h \rho  R_{\text{drum}}^2}}\right)$ as
\begin{align}
    \bar{\Omega}_1^2&=\frac{\omega _0^2 T_{\text{pre}}}{h \rho  R_{\text{drum}}^2}+\frac{3 \bar{a}_1^2 }{4 \alpha _1 h \rho  R_{\text{drum}}^4} \alpha _3 E_{2 D}-\frac{\bar{c}^2}{2 \rho ^2}-\sqrt{\frac{\bar{c}^4}{4 \rho ^4} -\frac{\bar{c}^2}{\rho ^2}\left(\frac{\omega _0^2 T_{\text{pre}}}{h \rho  R_{\text{drum}}^2}+\frac{3 \bar{a}_1^2 }{4 \alpha _1 h \rho  R_{\text{drum}}^4} \alpha _3 E_{2 D} \right)^2+\frac{\alpha _2^2 \bar{p}^2}{\alpha _1^2 h^2 \rho ^2 \bar{a}_1^2}} \label{Eq:correctFita1D},\\
    \bar{\Omega}_1^2&=\frac{\omega _0^2 T_{\text{pre}}}{h \rho  R_{\text{drum}}^2}+\frac{3  \bar{a}_2^2 }{4 \alpha _1 h \rho  R_{\text{drum}}^4}\alpha _3 E_{2 D}-\frac{\bar{c}^2}{2 \rho ^2}+\sqrt{\frac{\bar{c}^4}{4 \rho ^4} -\frac{\bar{c}^2}{\rho ^2}\left(\frac{\omega _0^2 T_{\text{pre}}}{h \rho  R_{\text{drum}}^2}+\frac{3 \alpha _3 \bar{a}_2^2 E_{2 D}}{4 \alpha _1 h \rho  R_{\text{drum}}^4} \alpha _3 E_{2 D}\right)^2+\frac{\alpha _2^2 \bar{p}^2}{\alpha _1^2 h^2 \rho ^2 \bar{a}_2^2}} \label{Eq:correctFita2D}.
\end{align}
As mentioned before, the unknowns in \cref{Eq:correctFita1D,Eq:correctFita2D} are  $\bc$ and $\alpha_3 E_{2D}$, with all other parameters known from the experiments.

\section{Conclusion\label{sec:Conclusion}}

In this article, we have presented direct numerical simulations of two coupled PDEs describing the oscillations of a membrane subject to a uniform, but oscillatory, load and linear damping. Using two different analytical approaches, Multiple Scales and Harmonic Balance, we obtained two theoretical predictions for the frequency--amplitude response; these were each analogous to previously derived results for the Duffing equation, but the analytical approaches used allowed us to detmerine the relevant prefactors in terms of the underlying physical properties of the system with no fitting parameters.

Our numerical simulations show a non-constant maximum compliance, which is often associated with nonlinear damping in experiments \cite[see][for example]{LifshitzReview}. Our simulations include only linear damping meaning that the inability of the Multiple Scales results to describe this feature of our numerical results is a deficiency of the method --- a deficiency that is not present in the results of Harmonic Balance. Indeed, the frequency--response curve determined by Harmonic Balance agrees well with the results of our numerical simulations up to, and including, the point of jump-down.

We finished by discussing how fits of the frequency--response curve could be made robust to the noise that is inevitably part of experiments, and which can lead to premature jump-down. Rather than focussing on the amplitude at the jump-down point alone, we suggest that taking the behaviour of the system either side of jump-down allows a fitting that is not confounded by premature jump-down. Using our numerical data, we demonstrate that performing the fit in this way provides fitted values of the key parameters that are closer to the values used in simulations and also provides a frequency--response curve that is much closer to the numerically generated data points.

We hope that the process we have outlined here may be generalized to provide reduced order models of, for example, NEMS/MEMS resonators subject to non-uniform loading, with prefactors in those models that are formally related to the underlying parameters.

\paragraph{Acknowledgement} We are grateful to Lincoln College, Oxford and IIT Hyderabad for funding, Prof.~Vyasarayani for discussions and Prof.~ Steeneken for providing his group's previous experimental data in figure~\ref{fig:ExptData} and comments on an earlier version of this manuscript.

\begin{appendix}
\setcounter{figure}{0}
\setcounter{equation}{0}
\setcounter{table}{0}
\renewcommand{\thefigure}{A.\arabic{figure}}
\renewcommand{\theequation}{A.\arabic{equation}}
\section{Validity of modelling assumptions}
\label{sec:SimplifyingAssumptions}
 
 The axisymmetric dynamic oscillations of a sheet must satisfy transverse force balance, which can be written as
\begin{equation}
\rho h \frac{\partial^2 \bZ}{\partial \bt^2}=\bp[\br,\bZ (\br,\bt),\bt]-B\nabla^4\bZ+\bm{\nabla}\cdot\left( \bm{\bSigma}\cdot\bm{\nabla}\bZ \right)
    \label{Eq:trans_force_balance}
\end{equation} and the in-plane equation of motion, which may be written as
\begin{equation}
\rho h \frac{\partial^2 \bm{\bu}}{\partial \bt^2}=\frac{1}{\br}\frac{\partial\left(\br\bSigma_{\br \br}\right)}{\partial \br} - \frac{\bSigma_{\bar{\theta} \bar{\theta}}}{\br}.
    \label{Eq:in_plane_force_balance}
\end{equation}
With the non-dimensionalization used throughout the main text, \cref{Eq:trans_force_balance} becomes
\begin{equation}
\frac{\partial^2 {\zeta}}{\partial {t}^2}={ p}({r},{t})-\bending \nabla^4{\zeta}+\bm{\nabla}\cdot\left( {\bm{\sigma}}\cdot\bm{\nabla}{\zeta} \right)
    \label{Eq:trans_force_balance_2}
\end{equation}
where
\begin{equation}
 \bending =\frac{B}{\Tpre \Rd^2}
    \label{Eq:epsilon}
\end{equation} is the reciprocal of the `bendability' \cite[][]{Davidovitch2011} of the membrane.
Now considering \cref{Eq:in_plane_force_balance}
\begin{equation}
 \frac{1}{{r}}\frac{\partial\left({r}{\sigma}_{r r}\right)}{\partial {r}} -\frac{{\sigma}_{\theta\theta}}{{r}}= \frac{\Tpre}{\ETwoD} \frac{\partial^2 {\bm{u}}}{\partial {t}^2}
\end{equation}
Thus, the dimensionless  time scale for evolution of the stress is $(\Tpre/\ETwoD)^{1/2}$, which corresponds to a dimensional time 
\begin{equation}
    \left(\frac{\Tpre}{\ETwoD} \frac{\rho h \Rd^2}{\Tpre}\right)^{1/2}=\frac{\Rd}{c},
    \label{eqn:SoundTime}
\end{equation}
where $c$ is the natural wave speed
\begin{equation}
c = \sqrt{\frac{\ETwoD}{\rho h}}.
\label{Eq:wave_speed}
\end{equation}

Our neglect of bending stiffness and in-plane inertia therefore require that both $\bending\ll1$ and $\omega\times\Rd/c\ll1$. By taking a typical Elastic Modulus value $E=400$~GPa, $h=3\times 10^{-10}\mathrm{~m}$, $\Rd=2.5\,\mu\mathrm{m}$, $\omega=10^7\mathrm{~Hz}$ \cite[all based on  figure 1b of][]{davidovikj_nonlinear_2017}, $\nu=0.3$ and $\Tpre\approx0.1\mathrm{~N\,m^{-1}}$  \cite[][]{lee_measurement_2008}, we get $\bending\approx6\times 10^{-5}$ and $\omega\Rd/c\approx 10^{-3}$; this suggests that both bending and dynamic effects can be neglected.

 \section{Details of the numerical scheme}
 \label{sec:AppendixNums}
 % \subsection{Description of scheme}
 %\setcounter{equation}{7}
 \renewcommand{\theequation}{B.\arabic{equation}}
 % \label{subsec:num_implementation}
We obtain the dynamics of the membrane by solving \cref{Eq:VForceBal_ND}--\cref{Eq:bcs_dimensionless}. 
To solve the PDE  \cref{Eq:VForceBal_ND}  numerically  by using the method of lines \cite[see][for example]{Schiesser2009}. We first  discretize the spatial derivative and convert \cref{Eq:VForceBal_ND} to a set of  ordinary differential equations. We do this by introducing a uniform mesh for $r$ on the interval $[0,1]$. The grid points on the mesh are given by $r_i=i/N$, where $0\leq i\leq N$ . We denote  the numerical values of $\zeta$, $\psi$ and $\psi'$ at the grid point $r_i$ by $\zeta_{i}$, $\psi_i$ and $\psi_i'$, respectively. The spatial derivatives of $\zeta$ in \cref{Eq:VForceBal_ND}, except at points $r=0$ and $r=1$, are approximated with second-order central differences. The spatial derivatives in $\zeta$ at $r=0$ and $r=1$ are approximated by second-order forward and backward differences, respectively. {(Note that $\psi'$ is calculated directly from our solution of the compatibility equation, and hence is not calculated via finite differences.)} Thus, for $i=1,2,\dots,N-1$, \cref{Eq:VForceBal_ND} becomes:  
\begin{equation}
    \frac{d^2 {\zeta_i}}{dt^2}+  \gamma\frac{d {\zeta_i}}{dt}={p}[{r}_i,\zeta_i,t]+\frac{\psi'_{i}}{r_i}\frac{{\zeta}_{i+1}-{\zeta}_{i-1}}{2 \Delta {r}}+\frac{\psi_{i}}{r_i}\frac{{\zeta}_{i+1}-2{\zeta}_i+{\zeta}_{i-1}}{\Delta {r}^2},
    \label{Eq:trans_force_balance_no_bending_discrete}
\end{equation}
and for $i=0$ and $N$, it becomes:
\begin{align}
    \frac{d^2 {\zeta_0}}{dt^2}+  \gamma\frac{d {\zeta_0}}{dt}&={p}[{r}_0,\zeta_0,t]+\frac{\psi'_{0}}{r_0}\frac{-3\zeta_0+4\zeta_1-\zeta_2}{2 \Delta {r}}\nonumber\\
    &+\frac{\psi_{0}}{r_0}\frac{2\zeta_0-5\zeta_1+4\zeta_2-\zeta_3}{\Delta {r}^2},    \label{Eq:trans_force_balance_no_bending_discrete_left}\\
    \frac{d^2 {\zeta_N}}{dt^2}+  \gamma\frac{d {\zeta_N}}{dt}&={p}[{r}_N,\zeta_N,t]+\frac{\psi'_{N}}{r_N}\frac{3\zeta_N-4\zeta_{N-1}+\zeta_{N-2}}{2 \Delta {r}} \nonumber\\
    &+\frac{\psi_{N}}{r_N}\frac{2\zeta_N-5\zeta_{N-1}+4\zeta_{N-2}-\zeta_{N-3}}{\Delta {r}^2},
    \label{Eq:trans_force_balance_no_bending_discrete_right}
\end{align}
respectively.  The boundary conditions, \cref{Eq:bcs_dimensionless}, become:
\begin{align}
    \zeta_0 &= \frac{4\zeta_1-\zeta_2}{3},   \quad &{\zeta}_{N}&=0     \label{Eq:bcs_dimensionless_zeta_dicrete},\\
    {\psi_0}&=0,  &{\psi}'_{N}-\nu \, {\psi}_N&=1-\nu.
    \label{Eq:bcs_dimensionless_psi_dicrete}
\end{align} 

The PDEs are thus recast as a set of ODEs. To write this set of ODEs in matrix form, we introduce the column vectors, $\bm{\zeta}$ and  $\bm{p}$, each of length $N-1$:
\begin{equation}
    \bm{\zeta} = \left({\zeta}_1,\dots,{\zeta}_{N-1}\right)^\mathrm{T},
\end{equation}
\begin{equation}
    \bm{p} = p\left(1,\dots,1\right)^\mathrm{T}.
\end{equation} The PDEs may then be written in the form
\begin{equation}
    \frac{d^2     \bm{\zeta}}{dt^2}=-\gamma \frac{d \bm{\zeta}}{dt}+\bm{p}+\left(\bm{S}_{rr}\bm{A}+\bm{S}_{\theta\theta}\bm{B}\right)\bm{\zeta}
    \label{Eq:system_of_equations}
\end{equation}  where the matrices $\bm{S}_{rr}$ and $\bm{S}_{\theta\theta}$ represent the effects of the in-plane stress and are defined as

%and matrices of size $(N-1) \times (N-1)$:
\begin{equation}
    \bm{S}_{rr} =  \begin{pmatrix}
\frac{\psi_1}{r_1}  &  &  \\
 &\ddots  &   \\
  &   & \frac{\psi_{N-1}}{r_{N-1}} \\
\end{pmatrix},  
\end{equation} and
\begin{equation}
    \bm{S_{\theta\theta}} =  \begin{pmatrix}
 \frac{\psi'_1}{r_1}  &  &  \\
 & \ddots  &     \\
& & \frac{\psi'_{N-1}}{r_{N-1}} \\
\end{pmatrix},  
\end{equation} while the matrices $\bm{A}$ and $\bm{B}$, defined by
\begin{equation}
    \bm{A} =\frac{1}{\Delta {r}^2}  \begin{pmatrix}
-2/3 & 2/3 &  &  & &  \\
1 & -2 & 1 &   &  &  \\
  & 1 & -2 & 1 &  &  \\
 &  &\ddots &\ddots &\ddots &   \\
 &  &  & 1 & -2 & 1  \\
  &  &  &  & 1 & -2 \\
\end{pmatrix},   
\end{equation}
\begin{equation}
    \bm{B} = \frac{1}{2 \Delta {r}} \begin{pmatrix}
-4/3 & 4/3 & 1 &   & &  & \\
-1 & 0 & 1 &   &  &  & \\
  & -1 & 0 & 1 &  &  & \\
 &  &\ddots &\ddots &\ddots &  &  \\
 &  &  & -1 & 0 & 1 & \\
  &  &  &  & -1 & 0 & 1\\
  &  &  &  &  & -1 & 0\\
\end{pmatrix},   
\end{equation} calculate the second and first derivatives of $\zeta$ with respect to $r$, respectively.

At each time step, the values of $\psi$ are calculated by solving \cref{Eq:compatibility_dimensionless} with the boundary conditions \cref{Eq:bcs_dimensionless_psi_dicrete}. The initial conditions are:
\begin{equation}
    \bm{\zeta}(0)=0, \quad\bm{\Dot{\zeta}}(0)=0. \label{Eq:initial_conditions} 
\end{equation} 
 
 We solve \cref{Eq:system_of_equations} by numerically integrating it in python  by using \texttt{solve\_ivp} routine with the initial conditions \cref{Eq:initial_conditions}. We solve for $\bm{\psi}$  by using \texttt{solve\_bvp} routine for \cref{Eq:compatibility_dimensionless} with the boundary conditions \cref{Eq:bcs_dimensionless_psi_dicrete}. To obtain the full response curve we do  forward and backward sweeps. The response branch with larger amplitudes and the jump-down point are found from the forward sweep. Similarly, the lower branch and the jump up point are from the backward sweep.

 Numerical error resulting from a coarse discretization can be minimized by taking a large $N$ but at the cost of increased computational time. We used a convergence study in which the value of $N$ is varied for each value of $\gamma$ used in the work to estimate this discretization error. We found that a value of  $N=50$ ensures reasonable computational time without compromising the accuracy of the results. We therefore use $N=50$ in the numerical solutions of  \cref{Eq:system_of_equations}--\cref{Eq:initial_conditions}  presented in the main text of this work.

\end{appendix}

%\bibliographystyle{elsarticle-num-names}
%\bibliography{refs.bib}

\end{document}